\documentclass[letterpaper,twocolumn,10pt]{article}
\usepackage{usenix2019_v3}

\usepackage{subfigure}
\usepackage{epstopdf}
\usepackage{amsmath} 
\usepackage{amsfonts} 
\usepackage{graphicx}
\usepackage{url}
\usepackage{hyperref}
\usepackage{booktabs}
\usepackage{mwe}
\usepackage{adjustbox}
\usepackage{enumitem}
\usepackage{xcolor}

\setlist{nolistsep}

\newcommand{\surf}{HPCCloud}
\newcommand{\google}{Google Cloud}
\newcommand{\amazon}{Amazon EC2}

\usepackage{booktabs}

\usepackage{multirow}
\usepackage{proofread}
\usepackage{xcolor}  
\definecolor{OwnAzure}{HTML}{336699}
\definecolor{OwnCerulean}{HTML}{CAE2FE}
\definecolor{OwnOliveGreen}{HTML}{556B2F}
\definecolor{KamPurple}{HTML}{907C97}
\usepackage[colorinlistoftodos,prependcaption,textsize=tiny]{todonotes}
\usepackage{xargs}                      
\newcommandx{\todolarge}[2][1=]{\todo[inline,size=\large,linecolor=OwnAzure,backgroundcolor=OwnCerulean,bordercolor=OwnAzure,#1]{#2}}
\newcommandx{\todoai}[2][1=]{\todo[inline,linecolor=OwnAzure,backgroundcolor=OwnCerulean,bordercolor=OwnAzure,#1]{#2}}

\newcommandx{\addref}[2][1=]
{\todo[inline,linecolor=blue,backgroundcolor=blue!50,bordercolor=blue,#1]{Add reference. #2}}
\newcommandx{\unsure}[2][1=]{\todo[linecolor=red,backgroundcolor=red!25,bordercolor=red,#1]{#2}}
\newcommandx{\change}[2][1=]{\todo[linecolor=blue,backgroundcolor=blue!25,bordercolor=blue,#1]{#2}}
\newcommandx{\info}[2][1=]{\todo[linecolor=OwnOliveGreen,backgroundcolor=OwnOliveGreen!25,bordercolor=OwnOliveGreen,#1]{#2}}
\newcommandx{\improvement}[2][1=]{\todo[linecolor=Plum,backgroundcolor=Plum!25,bordercolor=Plum,#1]{#2}}
\newcommandx{\nice}[2][1=]{\todo[linecolor=Plum,backgroundcolor=Plum!25,bordercolor=Plum,#1]{Nice! #2}}
\newcommandx{\thiswillnotshow}[2][1=]{\todo[disable,#1]{#2}}

\definecolor{darkred}{rgb}{0.5,0,0}
\definecolor{darkgreen}{rgb}{0,0.5,0}
\definecolor{darkblue}{rgb}{0,0,0.5}

\newcommand{\addition}[1]{{\color{black} #1}}

\usepackage{hyperref}

\begin{document}

\date{}

\title{\Large \bf Is Big Data Performance Reproducible in Modern Cloud Networks?}

\author{Alexandru Uta$^1$, Alexandru Custura$^1$, Dmitry Duplyakin$^2$, Ivo Jimenez$^3$, Jan Rellermeyer$^4$,\\ Carlos Maltzahn$^3$, Robert Ricci$^2$, Alexandru Iosup$^1$\\
\small {\em  $^1$VU Amsterdam \quad
          $^2$University of Utah \quad
          $^3$University of California, Santa Cruz \quad
          $^4$TU Delft } \\ [2mm]
}

\maketitle

\begin{abstract}
Performance variability has been acknowledged as a problem for over a decade by cloud practitioners and performance engineers. Yet, our survey of top systems conferences reveals that the research community regularly disregards variability when running experiments in the cloud. Focusing on networks, we assess the impact of variability on cloud-based big-data  workloads by gathering traces from mainstream commercial clouds and private research clouds. Our data collection consists of millions of datapoints gathered while transferring over 9 petabytes of data. 
We characterize the network variability present in our data and show that, even though commercial cloud providers implement mechanisms for quality-of-service enforcement, variability still occurs, and is even exacerbated by such mechanisms and service provider policies. 
We show how big-data workloads suffer from significant slowdowns and lack predictability and replicability, even 
when state-of-the-art experimentation techniques are used.
We provide guidelines for practitioners to reduce the volatility of big data performance, making experiments more repeatable.
\end{abstract}

\section{Introduction}

Performance variability~\cite{maricq2018taming,cao2017performance} in the cloud is well-known, and has been studied since the early days~\cite{schad2010runtime,iosup2010performance,ballani2011towards} of cloud computing. Cloud performance variability impacts not only operational concerns, such as cost and predictability~\cite{casale2013modelling,leitner2016patterns}, but also reproducible experiment design~\cite{maricq2018taming,hoefler2015scientific,blackburn2016truth,abedi2017conducting}. 
 Big data is now highly embedded in the cloud: Hadoop~\cite{white2012hadoop} or Spark~\cite{zaharia2012resilient} processing engines have been deployed for many years on on-demand resources. 
 One key issue when running big data workloads in the cloud is that, due to the multi-tenant nature of clouds, applications see performance effects from other tenants, and are thus susceptible to performance variability, including on the network. Even though recent evidence~\cite{ousterhout2015making} suggests that there are limited potential gains from \emph{speeding up} the network, it is still the case that variable network performance can \emph{slow down} big data systems and introduce volatility that makes it more difficult to draw reliable scientific conclusions.
 
Although cloud performance variability has been thoroughly studied, the resulting work has mostly been in the context of optimizing tail latency~\cite{dean2013tail}, with the aim of providing more consistent application-level performance~\cite{chaimov2016scaling,suresh2015c3,grosvenor2015queues,ghit2015reducing}. This is subtly---but importantly---different from understanding the ways that fine-grained, resource-level variability affects the \emph{performance evaluation} of these systems. Application-level effects are especially elusive for complex applications, such as big data, which are not bottlenecked on a specific resource for their entire runtime. As a result, it is difficult for experimenters to understand how to design experiments that lead to reliable conclusions about application performance under network variability conditions.

Modern cloud data centers increasingly rely on software-defined networking to offer flows between VMs with reliable and predictable performance~\cite{mogul2012we}. 
While modern cloud networks generally promise isolation and predictability~\cite{ballani2011towards,guo2010secondnet}, in this paper we uncover that they rarely achieve stable performance. Even the mechanisms and policies employed by cloud providers for offering quality of service (QoS) and fairness can result in non-trivial interactions with the user applications, which leads to performance variability. 


Although scientists are generally aware of the relationship between repeated experiments and increased confidence in results, the specific strength of these effects, their underlying causes, and methods for improving experiment designs have not been carefully studied in the context of performance experiments run in clouds. Variability has a significant impact on sound experiment design and result reporting~\cite{hoefler2015scientific}. In the presence of variability, large numbers of experiment repetitions must be performed to achieve tight confidence intervals~\cite{maricq2018taming}. Although practitioners and performance engineers acknowledge this phenomenon~\cite{schad2010runtime,iosup2010performance,ballani2011towards}, in practice these effects are frequently disregarded in performance studies.

Building on our vision~\cite{iosup2018massivizing}, we challenge the current state-of-practice in cloud-based systems experimentation and advocate for sound experiment design and result reporting. We show that, due to performance variability, flawed cloud-based experimentation could lead to inaccurate or even wrong conclusions. We show, in-depth, the performance implications of network variability when running big data workloads. The interplay between underlying resources and applications is complex, and leads to non-trivial performance behavior. 

To characterize such interactions, we run state-of-the-art, real-world applications using Apache Spark~\cite{armbrust2015spark}. We run big-data workloads either directly on real-world mainstream clouds, or by emulating the network behavior of such clouds. Our results show that variability highly impacts not only performance, but also credible and reproducible experimentation. 

Addressing cloud users, performance engineers, and system designers, we examine the implications of \emph{network} variability on big data, and present our main findings and contributions:
\begin{enumerate}[leftmargin=0cm,itemindent=.5cm,labelwidth=\itemindent,labelsep=0cm,align=left]
    \item \textbf{Lack of sound experimentation:} Many articles in the literature that present cloud-based experiments are either under-specified (i.e., do not report statistical measures), or run inconclusive numbers of experiment repetitions (Section~\ref{sec:litsurv}).
    \item \textbf{Variability in modern cloud networks:} We conduct and analyze measurements of public and private cloud providers, and characterize the level of variability and identify specific sources of variability (Section~\ref{sec:netw_var}).
    \item \textbf{Network variability impact on application performance reproducibility:} Low-level network variability can have significant effects on application performance, and can cause violations of assumptions commonly used in performance modeling (such that experiment runs are independent and identically distributed) (Section~\ref{sec:mechanisms_effect}).
    \item \textbf{Strategies for running reproducible experiments:} Given our measurement and experience with application-level benchmarks, we make recommendations for improving the reliability and reproducibility of experiments (Section~\ref{sec:protocol}).
\end{enumerate}

\begin{figure}[t]
    \vspace*{-0.2cm}
    \centering
    \includegraphics[width=0.99\columnwidth]{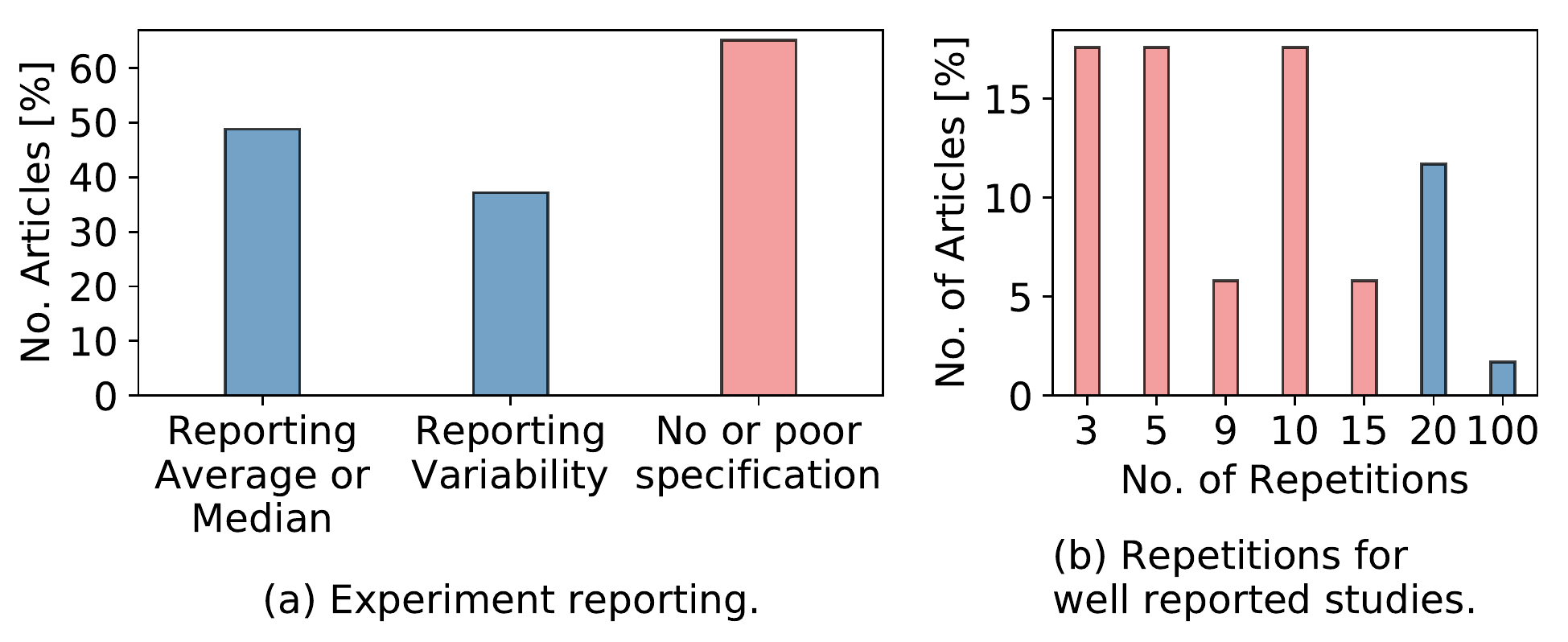}
    \vspace*{-0.35cm}
    \caption{State-of-practice in big data articles with cloud experiments: (a) Aspects reported about experiments. Bars represent aspects that are not mutually exclusive, thus total can exceed 100\%. (b) Number of experiment repetitions performed for the properly specified articles.}
    \label{fig:underspecified_articles}
    \vspace*{-0.4cm}
\end{figure}

\section{Is Cloud Variability Disregarded?}\label{sec:litsurv}

We perform a systematic survey to uncover whether and how researchers and practitioners take cloud performance variability into account when running experiments. Our findings are depicted in Figure~\ref{fig:underspecified_articles} and summarized as follows:
\begin{description}[noitemsep,nosep,leftmargin=-0.03cm]
\item[Finding 2.1] Cloud performance variability is largely disregarded when researchers evaluate and prototype distributed systems, or when comparing established systems.
\item[F2.2] Most cloud performance studies are severely under-specified. Most studies: (i) do not report what performance measures are reported (i.e., mean, median); (ii) do not report minimal statistical data (i.e., standard deviation, quartiles); (iii) do not report the number of repetitions of an experiment.
\item[F2.3] Most cloud performance evaluations are poorly designed: a large majority of such studies only perform small numbers of experiment repetitions (i.e., 3-10 trials), 
and do not assess variability or confidence.
\end{description}

\begin{table}[]
\caption{Parameters for the performance variability systematic survey. Further, we manually select only the articles with empirical evaluations performed using clouds.}
\resizebox{0.98\columnwidth}{!}{
\begin{tabular}{@{}lll@{}}
\toprule
\textbf{Venues}                                                                & \textbf{Keywords}                                                                                                                           & \textbf{Years}       \\ \midrule
\begin{tabular}[c]{@{}l@{}} NSDI, OSDI\\ SOSP, SC\end{tabular} & \begin{tabular}[c]{@{}l@{}}big data, streaming, Hadoop,\\ MapReduce, Spark, data storage\\ graph processing, data analytics\end{tabular} & 2008 - 2018 \\ \bottomrule
\end{tabular}
}
\label{tab:litsurv}
\end{table}


\begin{table}[t]
\caption{Survey process. Initial filtering done automatically by keywords, then manually for cloud-based experiments. Resulting subset is significant and highly-cited.}
\resizebox{0.98\columnwidth}{!}{
\begin{tabular}{llll}
\hline
{  \textbf{\begin{tabular}[c]{@{}l@{}}Articles\\ Total\end{tabular}}} & {  \textbf{\begin{tabular}[c]{@{}l@{}}Filtered\\Automatically\\ by Keywords\end{tabular}}} & {  \textbf{\begin{tabular}[c]{@{}l@{}}Filtered \\ Manually\\ for Cloud \\ Experiments\end{tabular}}} & {  \textbf{\begin{tabular}[c]{@{}l@{}}Citations for\\ selected\\  44 articles\end{tabular}}} \\ \hline
{  1,867}                                                             & {  138}                                                                                       & {  \begin{tabular}[c]{@{}l@{}}44 (15 NSDI, 7 OSDI,\\ 7 SOSP, 15 SC)\end{tabular}}                  & {  11,203}                                                                                \\ \hline
\end{tabular}
}
\label{tab:litsurv_process}
\vspace*{-0.1cm}
\end{table}

Over the last decade, big data processing platforms and applications have been co-evolving with the cloud. This allowed researchers and practitioners to develop, deploy, and evaluate their applications and systems on various virtualized infrastructures: public, private, and hybrid. 
There is much evidence that clouds suffer from performance variability~\cite{iosup2010performance,ballani2011towards,maricq2018taming,cao2017performance}. It is therefore intuitive to ask whether or not practitioners and system designers take variability into account when designing experiments, or when building systems. To answer these questions, we performed a systematic literature survey covering prominent conferences in the field: NSDI~\cite{DBLP:conf/nsdi/2019}, OSDI~\cite{DBLP:conf/osdi/2018}, SOSP~\cite{DBLP:conf/sosp/2017}, and SC~\cite{DBLP:conf/sc/2018}. 

\textbf{Survey Methodology:} Table~\ref{tab:litsurv} shows the parameters of our survey, and \addition{Table~\ref{tab:litsurv_process} presents our survey process in-depth: (1) we started with all articles published in the aforementioned venues; (2) selected automatically a subset, based on string matching our query on keywords, title, and abstract; (3) we manually selected the articles in which the experiments were performed on a public cloud. The 44 selected articles are highly influential, having been cited \textbf{11,203} times so far\footnote{according to Google Scholar on May 20, 2019}.} 

The criteria we looked for when analyzing such articles are the following: (i) reporting average or median metrics over a number of experiments; (ii) reporting variability (such as standard deviation or percentiles) or confidence (such as confidence intervals); (iii) reporting the number of times an experiment was repeated. These are all critical criteria for determining whether a study’s conclusions may be irreproducible, or worse, not fully supported by the evidence (i.e., flawed). All manual filtering was performed by two separate reviewers, and we applied Cohen's Kappa coefficient~\cite{cohen1960coefficient} for each category presented in Figure~\ref{fig:underspecified_articles}a: reporting average or median, statistics, and poor specification. Our Kappa scores, for each category, were 0.95, 0.81, and 0.85, respectively. For Cohen's Kappa, values larger than 0.8 show that \emph{almost perfect agreement} has been achieved by the reviewers~\cite{viera2005understanding}. Our findings are detailed in the following paragraphs.

\textbf{Survey Results:} \emph{The systems community centered around cloud computing and big data gravely disconsiders performance variability when performing empirical evaluations in the cloud.} Figure~\ref{fig:underspecified_articles} shows the results of our survey. \addition{Out of the two reviewer's scores, we plot the lower scores, i.e., ones that are more favorable to the articles.} We found that over 60\% of the surveyed articles are severely under-specified (i.e., the authors do not mention how many times they repeated the experiments or even what numbers they are reporting, e.g., averages, medians); a subset of the articles report averages or medians, but out of those, only 37\% report variance or confidence (i.e., error-bars, percentiles). We further found that most articles that do report repetitions perform only 3, 5 or 10 repetitions of the experiments. The reason for such practices might be that experimenters are still more used to evaluating software in controlled 
environments---what is true in controlled environments often does not hold in clouds.

Moreover, 76\% of the properly specified studies use 
no more than 15 repetitions. Coupled with the effects of cloud variability, such experiment design practices could lead to wrong or ambiguous conclusions, as we show next. 

\begin{figure}[t]
\centering	
\includegraphics[width=0.82\linewidth]{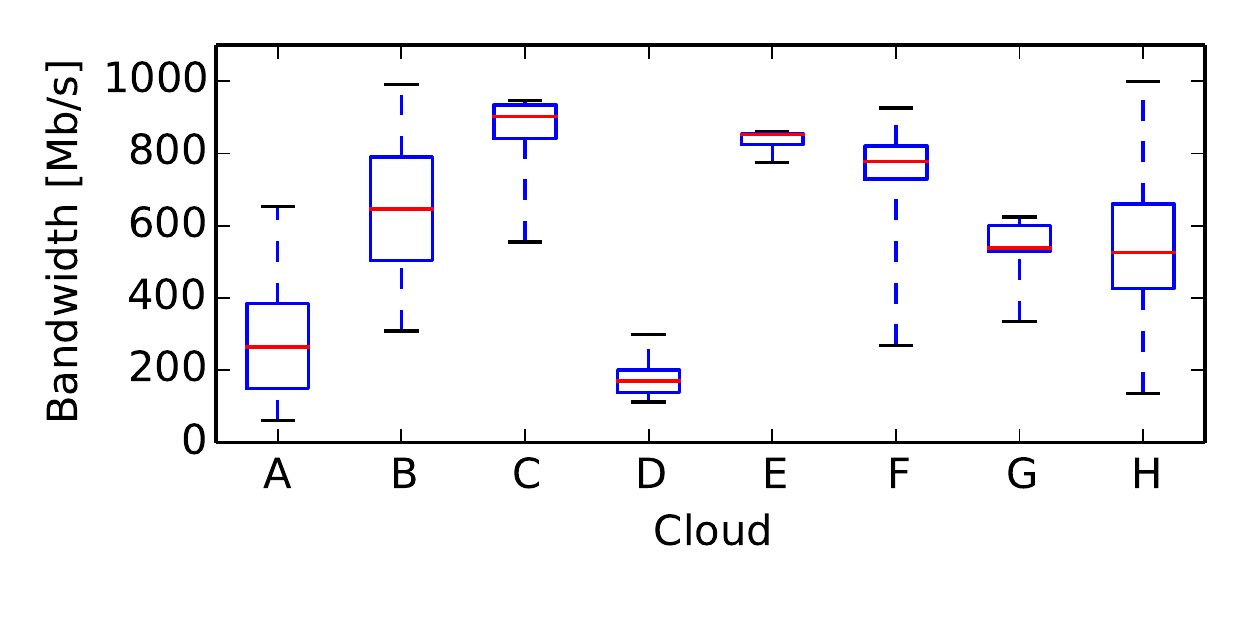}
\vspace{-0.5cm}
\caption{Bandwidth distributions for eight real-world clouds. Box-and-whiskers plots show the 1st, 25th, 50th, 75th, and 99th percentiles. (Distributions derived from the study~\cite{ballani2011towards} conducted by Ballani et al.)}
\label{fig:bw_distribution}
\vspace*{-0.2cm}
\end{figure}

\begin{figure}[t]
    \centering
    \begin{subfigure}{\small{(a) Medians for HiBench-KMeans}}
       \includegraphics[width=0.92\columnwidth]{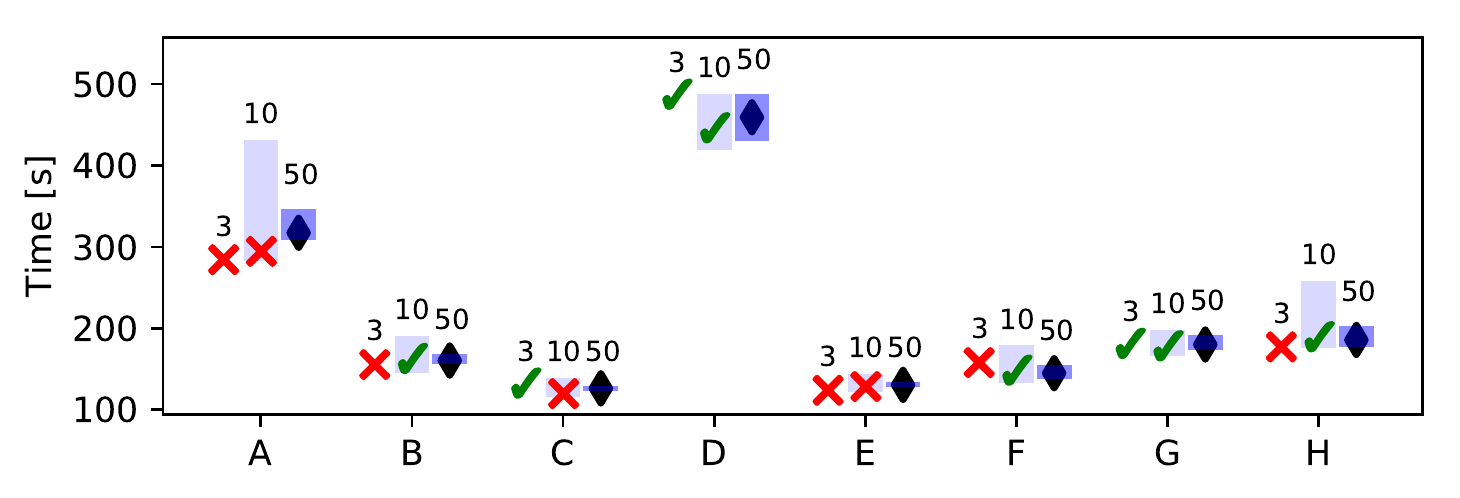}
           \vspace{-0.2cm}
       \label{fig:ci-hibench} 
    \end{subfigure}
    \begin{subfigure}{\small{(b) 90th percentile for TPC-DS Q68}}
       \includegraphics[width=0.92\columnwidth]{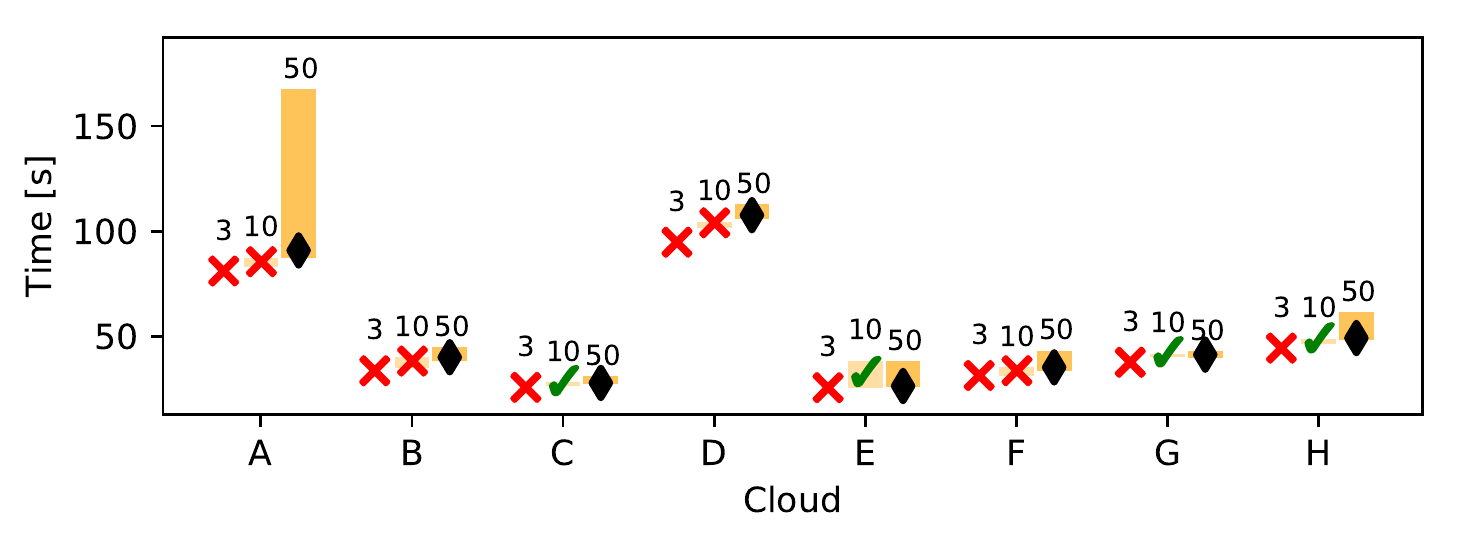}
       \label{fig:ci-tpcds}
    \end{subfigure}
    \vspace*{-0.5cm}
    \caption{Medians and 90th percentiles for K-Means (a) and TPC-DS Q68 (b). Estimates are shown along with their 95\% confidence intervals (CIs) for performance measurements under the A-H distributions. $\diamond$ depicts estimates 50-runs. Judged by the 50-run CIs we consider \emph{gold standard}, accurate estimates (inside those CIs) are \checkmark; inaccurate estimates (outside those CIs) are $\times$ for 3- and 10-run sets.}
    \label{fig:poor_experimentation}
    \vspace*{-0.4cm}
\end{figure}

\subsection{How credible are experiments with few repetitions?}

Experiments with few repetitions run the risk of reporting inaccurate results; the higher the variability, the greater the risk that a low-repetition experiment's results are unreliable. We use application-level benchmarks to show how the bandwidth distributions found by Ballani et al.~\cite{ballani2011towards} for eight real-world clouds---shown in Figure~\ref{fig:bw_distribution}---do affect findings in practice.


We emulate the behavior of the eight clouds presented in Figure~\ref{fig:bw_distribution}, which were contemporary with most articles found in our survey. In a private Spark~\cite{armbrust2015spark} cluster of 16 machines, we limit the bandwidth achieved by machines according to distributions $A-H$. We uniformly sample bandwidth values from these distributions every $x \in \{5, 50\}$ seconds.
We used 50 experiment repetitions as our “gold standard,” and compared them to the 3- and 10-repetitions commonly found in our literature survey.

\textbf{Emulation Results:} 
We found that experiments with few repetitions often produced medians that are \emph{outside of} the 95\% confidence intervals (CIs) for larger experiment sequences. The 95\% CIs for medians represent the ranges in which we would find the true medians with the 95\% probability, if we were able to run infinite repetitions. Thus, when the low-repetition medians lie outside of the high-repetition CIs, there is a 95\% probability that the former are inaccurate.
This can be seen in Figure~\ref{fig:poor_experimentation}, which plots estimates of 95\% nonparametric (asymmetric) CIs~\cite{Boudec2011Performance} for experiments using bandwidth distributions $A-H$ from Figure~\ref{fig:bw_distribution}. For each bandwidth distribution, we show the medians and CIs for 3-, 5-, and 50-repetition experiments.\footnote{Three repetitions are insufficient to calculate CIs; we include medians for this experiment setup because it is representative of what is often found in the literature.} The median for the “gold standard” experiment is marked with a diamond; medians for lower-repetition experiments are shown with an “X” if outside the gold-standard 95\% CI, or a check-mark if within it.

The top of Figure~\ref{fig:poor_experimentation} (part (a)) shows our estimates of \emph{medians} for the K-Means application from HiBench~\cite{huang2010hibench}. Of the eight cloud bandwidth distributions, the 3-run median falls outside of the gold-standard CI for six of them (75\%), and the 10-run median for three (38\%).
The bottom half of Figure~\ref{fig:poor_experimentation} (part (b)) shows the same type of analysis, but this time, for tail performance~\cite{dean2013tail} instead of the median.
To obtain these results, 
we used TPC-DS~\cite{nambiar2006making} Query-68 measurements and the method from
Le~Boudec~\cite{Boudec2011Performance} to calculate nonparametric estimates for the 90th percentile performance, as well as their confidence bounds.
As can be seen in this figure, it is even more difficult to get robust \emph{tail} performance estimates. 



\textbf{Emulation Methodology:} The quartiles in Ballani's study (Figure~\ref{fig:bw_distribution}) give us only a rough idea about the probability densities and there is uncertainty about fluctuations, as there is no data about sample-to-sample variability. Considering that the referenced studies reveal no autocovariance information, we are left with using the available information to sample bandwidth uniformly. Regarding the sampling rate, we found the following: (1) As shown in Section~\ref{sec:netw_var} two out of the three clouds we measured exhibits significant sample-to-sample variability on the order of tens of seconds; (2) The cases F-G from Ballani’s study support fine sampling rates: variability at sub-second scales~\cite{wang2010impact} and at the 20s intervals~\cite{farley2012more} is significant. Therefore, we sample at relatively fine-grained intervals: 5s for Figure~\ref{fig:poor_experimentation}(a), and 50s for Figure~\ref{fig:poor_experimentation}(b). Furthermore, sampling at these two different rates shows that benchmark volatility is not dependent on the sampling rate, but rather on the distribution itself.

\section{How Variable Are Cloud Networks?}\label{sec:netw_var}


We now gather and analyze network variability data for three different clouds:
two large-scale commercial clouds, and a smaller-scale private research cloud.
Our main findings can be summarized as follows:
\begin{description}[noitemsep,nosep,leftmargin=-0.03cm]
    
    \item[F3.1] Commercial clouds implement various mechanisms and policies for network performance QoS enforcement, and these policies are opaque to users and vary over time. We found (i) token-bucket approaches, where bandwidth is cut by an order of magnitude after several minutes of transfer; (ii) a per-core bandwidth QoS, prioritizing heavy flows; (iii) instance types that, when created repeatedly, are given different bandwidth policies unpredictably.
  
    \item[F3.2] Private clouds can exhibit more variability than public commercial clouds. Such systems are orders of magnitude smaller than public clouds (in both resources and clients), meaning that when competing traffic does occur, there is less statistical multiplexing to ``smooth'' out variation.
    
    \item[F3.3] Base latency levels can vary by a factor of almost $10x$ between clouds, and implementation choices in the cloud's virtual network layer can cause latency variations over two orders of magnitude depending on the details of the application.
    
\end{description}

\subsection{Bandwidth}\label{subsec:var_postgbit_networks}

We run our bandwidth measurements in two prominent commercial clouds, \amazon{} (us-east region) and \google{} (us-east region), and one private research cloud, \surf{}\footnote{\url{https://userinfo.surfsara.nl/systems/hpc-cloud}}. Table~\ref{tab:experiment_bw_variability} summarizes our experiments. In the interest of space, in this paper we focus on three experiments; all data we collected is available in our repository~\cite{our_traces}. We collected the data between October 2018 and February 2019. In total, we have over 21 weeks of nearly-continuous data transfers, which amount for over 1 million datapoints and over 9 petabytes of transferred data.  


The Amazon instances we chose are typical instance types that a cloud-based big data company offers to its customers~\cite{databricks_instances}, and these instances have AWS's “enhanced networking capabilities”~\cite{aws_enhanced_networking}. On \google{} (GCE), we chose the instance types that were as close as possible (though not identical) to the \amazon{} offerings. \surf{} offered a more limited set of instance types.
We limit our study to this set of cloud resources and their network offerings, as big data frameworks are not equipped to make use of more advanced networking features (i.e., InfiniBand or even 20+ Gbit Ethernet), as they are generally designed for commodity hardware. Moreover, vanilla Spark deployments, using typical data formats such as Parquet or Avro, are not able to routinely exploit links faster than 10\,Gbps, unless significant optimization is performed~\cite{trivedi2018albis}. 
Therefore, the results we present in this article are highly likely to occur in real-world scenarios.

\begin{table}[tp]
\caption{Experiment summary for determining performance variability in modern cloud networks. Experiments marked with a star (*) are presented in depth in this article. Due to space limitations, we release the other data in our repository~\cite{our_traces}. All \amazon{} instance types are typical offerings of a big data processing company~\cite{databricks_instances}.}
\resizebox{0.98\columnwidth}{!}{
\begin{tabular}{@{}llrlcr@{}}
\toprule
\multicolumn{1}{l}{\textbf{Cloud}} & \multicolumn{1}{c}{\textbf{\begin{tabular}[c]{@{}c@{}}Instance\\  Type\end{tabular}}} & \multicolumn{1}{c}{\textbf{\begin{tabular}[c]{@{}c@{}}QoS\\(Gbps)\end{tabular}}}  & \multicolumn{1}{c}{\textbf{\begin{tabular}[c]{@{}c@{}}Exp. \\ Duration\end{tabular}}} & \multicolumn{1}{c}{\textbf{\begin{tabular}[c]{@{}c@{}}Exhibits \\ Variability\end{tabular}}} &
\multicolumn{1}{c}{\textbf{\begin{tabular}[c]{@{}c@{}}Cost \\ (\$)\end{tabular}}}\\ \midrule
*Amazon                            & c5.XL                                                                             & $\leq$ 10                      & 3 weeks                               & Yes   & 171                                                                                         \\
Amazon                             & m5.XL                                                                             & $\leq$ 10                      & 3 weeks                               & Yes  & 193                                                                                           \\
Amazon                             & c5.9XL                                                                            & 10                           & 1 day                                 & Yes & 73                                                                                            \\
Amazon                             & m4.16XL                                                                           & 20                           & 1 day                                 & Yes & 153                                                                                            \\
Google                             & 1 core                                                                                & 2                            & 3 weeks                               & Yes & 34                                                                                            \\
Google                             & 2 core                                                                                & 4                            & 3 weeks                               & Yes & 67                                                                                            \\
Google                             & 4 core                                                                                & 8                            & 3 weeks                               & Yes & 135                                                                                            \\
*Google                            & 8 core                                                                                & 16                           & 3 weeks                               & Yes & 269                                                                                            \\
HPCCloud                               & 2 core                                                                                & N/A                              & 1 week                                & Yes & N/A                                                                                            \\
HPCCloud                               & 4 core                                                                                & N/A                              & 1 week                                & Yes & N/A                                                                                            \\
*HPCCloud                              & 8 core                                                                                & N/A                              & 1 week                                & Yes & N/A                                                                                            \\ \bottomrule
\end{tabular}
}
\label{tab:experiment_bw_variability}
\vspace*{-0.3cm}
\end{table}

In the studied clouds, for each pair of VMs of similar instance types, we measured bandwidth continuously for one week. Since big data workloads have different network access patterns, we tested multiple scenarios:
\begin{itemize}
    \item \textbf{full-speed} - continuously transferring data, and summarizing performability metrics (bandwidth, retransmissions, CPU load etc.) every 10 seconds;
    \item \textbf{10-30} - transfer data 10 seconds, wait 30 seconds; 
    \item \textbf{5-30} - transfer data 5 seconds, wait 30 seconds.
\end{itemize}
The first transmission regime models highly network intensive applications, such as long-running batch processing, or streaming. The last two modes mimic short-lived analytics queries, such as TPC-H, or TPC-DS. 


\begin{figure}[t]
    \centering
    \includegraphics[width=\columnwidth]{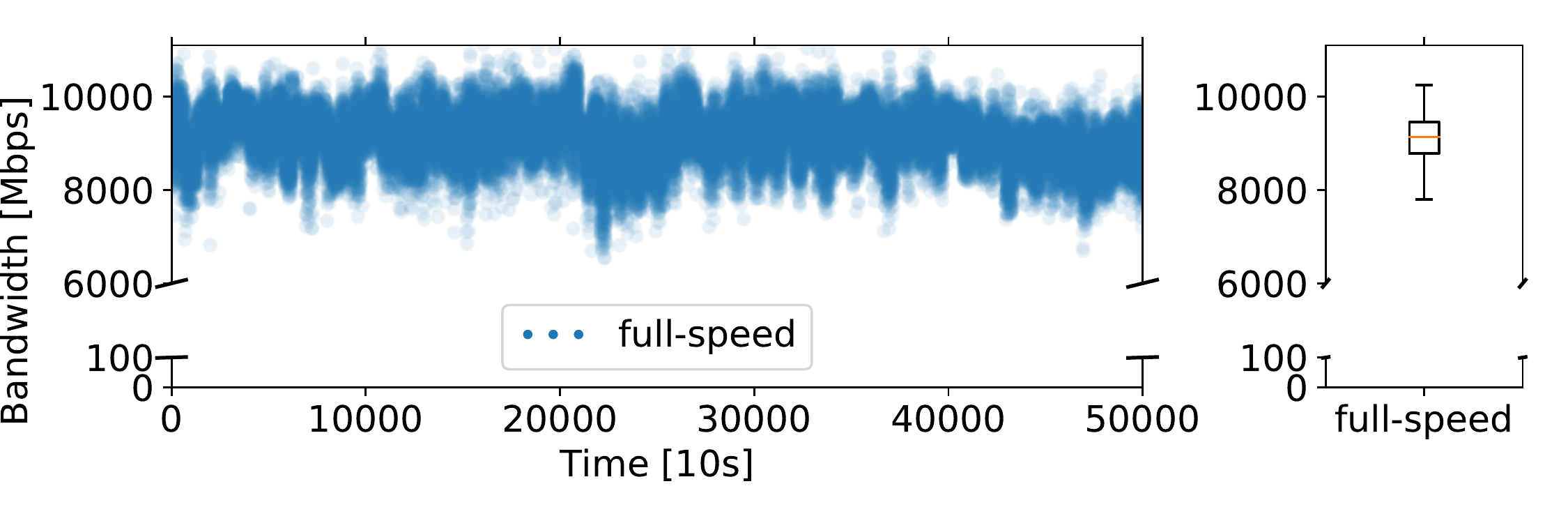}
    \vspace*{-0.7cm}
    \caption{Variable network bandwidth performance in the \surf{} (left); the statistical performance distribution, plotted as an IQR box; the whiskers represent 1st and 99th percentiles (right). Duration: a week of continuous experimentation; each point is average over 10 seconds.}
    \label{fig:perf_var_surf}
    \vspace*{-0.35cm}
\end{figure}

\textbf{\surf{}.} Small-scale (i.e., up to 100 physical machines and several hundred users) private (research) clouds do not use mechanisms to enforce network QoS. We measured the network performance variability between pairs of VMs, each having 8 cores. Figure~\ref{fig:perf_var_surf} plots the results. We show our measurements only for "full-speed" (i.e., continuous communication) because our other experiments 
show similar behavior.
We observe that the network bandwidth shows high variability, ranging from 7.7Gbps to 10.4Gbps. 

\begin{figure}[t]
    \centering
    \includegraphics[width=\columnwidth]{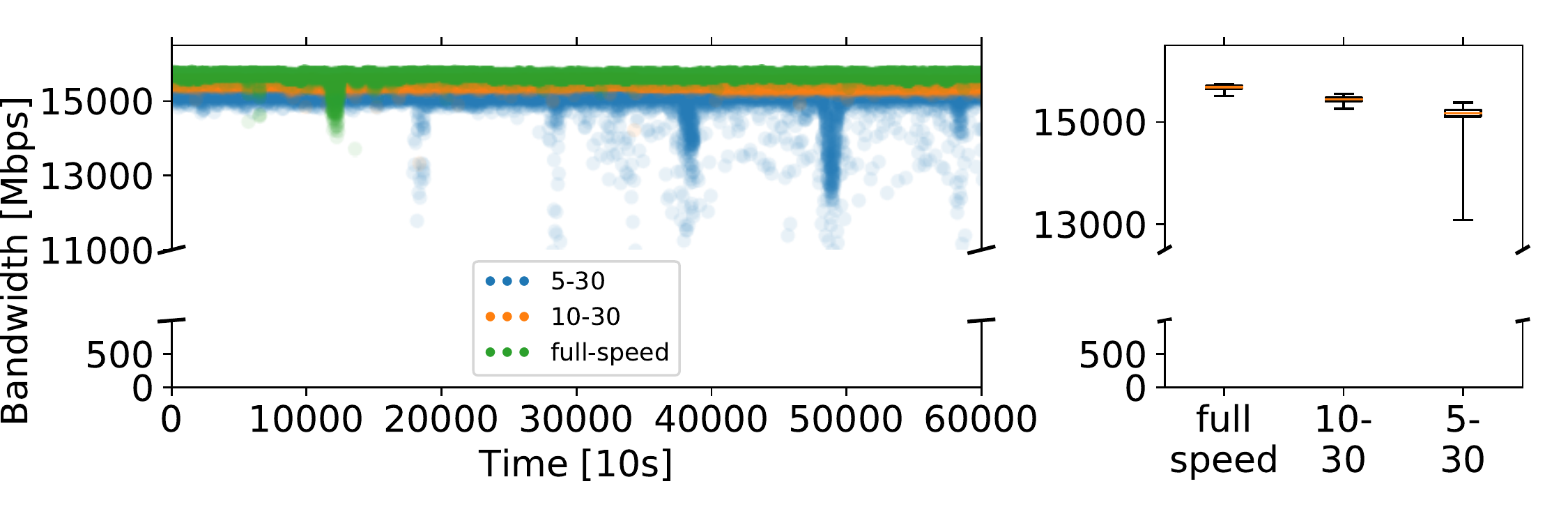}
    \vspace*{-0.85cm}
    \caption{Variable network bandwidth performance in the \google{} (left), and the statistical performance distribution, plotted as an IQR box, where the whiskers are 1st and 99th percentiles (right). The duration is a week of continuous experimentation, each point is an average over 10 seconds.}
    \label{fig:perf_var_google}
    \vspace*{-0.35cm}
\end{figure}

\textbf{\google{}.} GCE states that they enforce network bandwidth QoS by guaranteeing a "per-core" amount of bandwidth. Our measurements fall close to the QoS reported by the provider. Figure~\ref{fig:perf_var_google} plots the result of the experiments performed in the \google{}. We notice that the access pattern affects variability to a greater degree than in other clouds: longer streams exhibit low variability, and better overall performance: \emph{full-speed} achieves stable and high performance, while \emph{5-30} has a fairly long tail. This could be due to the design of the \google{} network, where idle flows use dedicated gateways for routing through the virtual network~\cite{dalton2018andromeda}. We observe that the network bandwidth varies significantly, depending on access patterns, between 13Gbps and 15.8Gbps.

\begin{figure}[h]
    \includegraphics[width=\columnwidth]{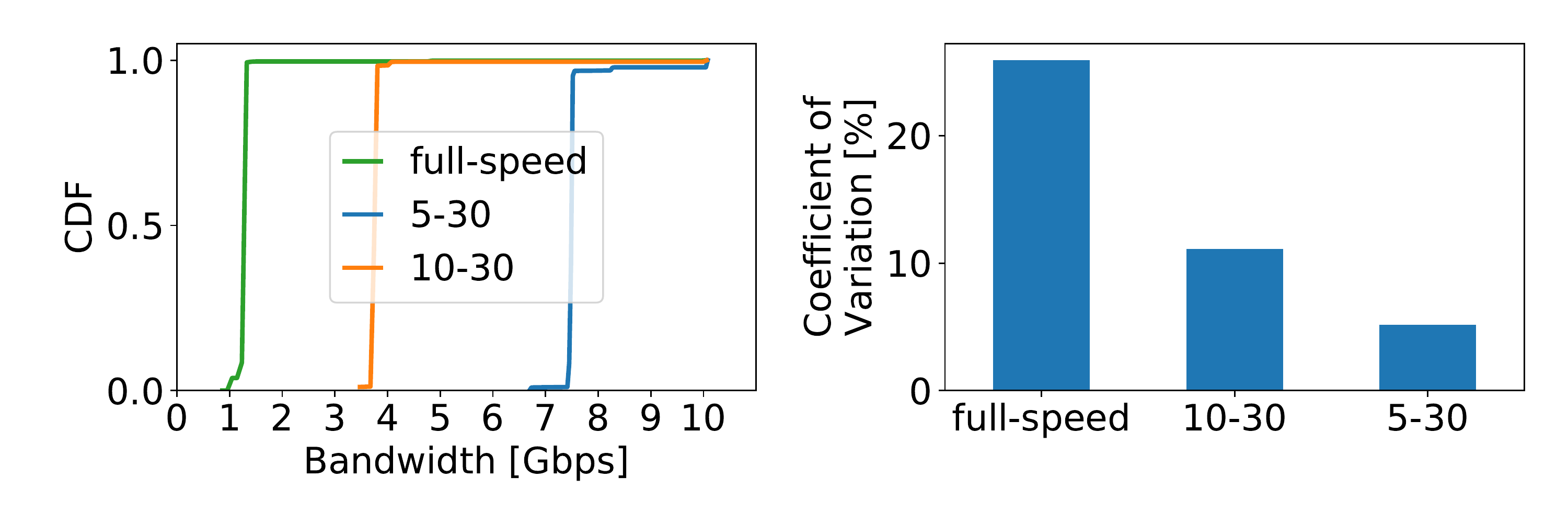}
    \vspace*{-0.65cm}
    \caption{Variable network bandwidth performance in the \amazon{}, plotted as an empirical cumulative distribution (left), barplot of the coefficient of variation (right). The duration is a week of continuous experimentation, each data point representing an average over 10 seconds.}
    \label{fig:perf_var_aws}
    \vspace*{-0.25cm}
\end{figure}

\textbf{\amazon{}.} We discover the opposite behavior in EC2: heavier streams achieve less performance and more variability compared to lighter (shorter) streams, as shown in Figure~\ref{fig:perf_var_aws}. Considering the large performance differences between these experiments, we plot our measurements as a CDF and a barplot of coefficient of variation to improve visibility. There are approximately 3x and 7x slowdowns between \emph{10-30} and \emph{5-30} and \emph{full-speed}, respectively. The achieved bandwidth varies between 1Gbps and 10Gbps. We look into the causes for this behavior later in the paper.

\begin{figure}[t]
    \centering
    \includegraphics[width=0.99\columnwidth]{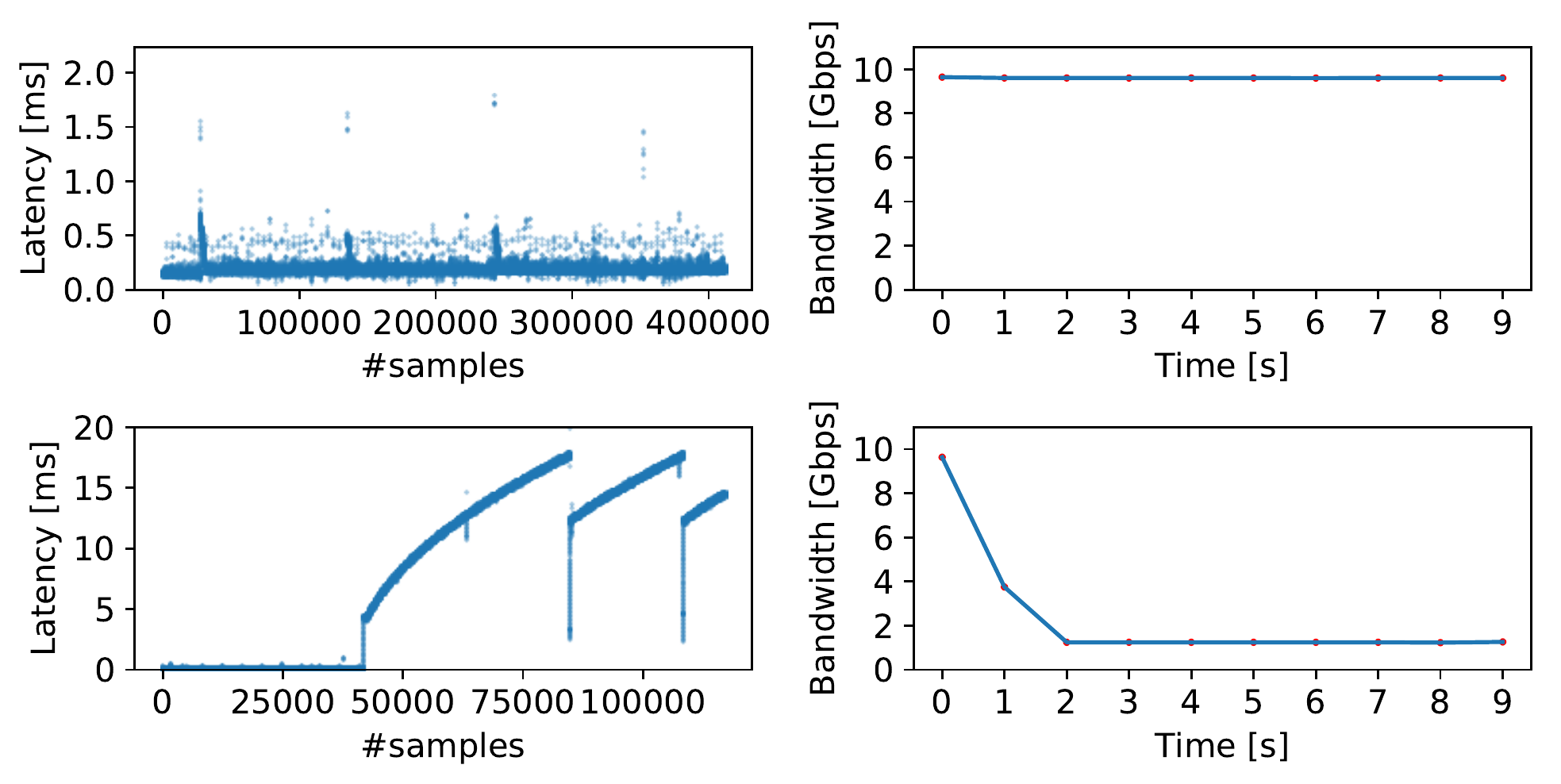}
    \vspace*{-0.2cm}
    \caption{Example of observed \amazon{} latency for a 10-second TCP sample on \emph{c5.xlarge}. Left: RTT latency for TCP packets. Right: achieved \emph{iperf} bandwidth. Top: regular \amazon{} behavior. Bottom: latency behavior when a drop in bandwidth occurs.}
    \label{fig:aws_latency}
    \vspace*{-0.2cm}
\end{figure}

\begin{figure}[t]
    \centering
    \includegraphics[width=0.99\columnwidth]{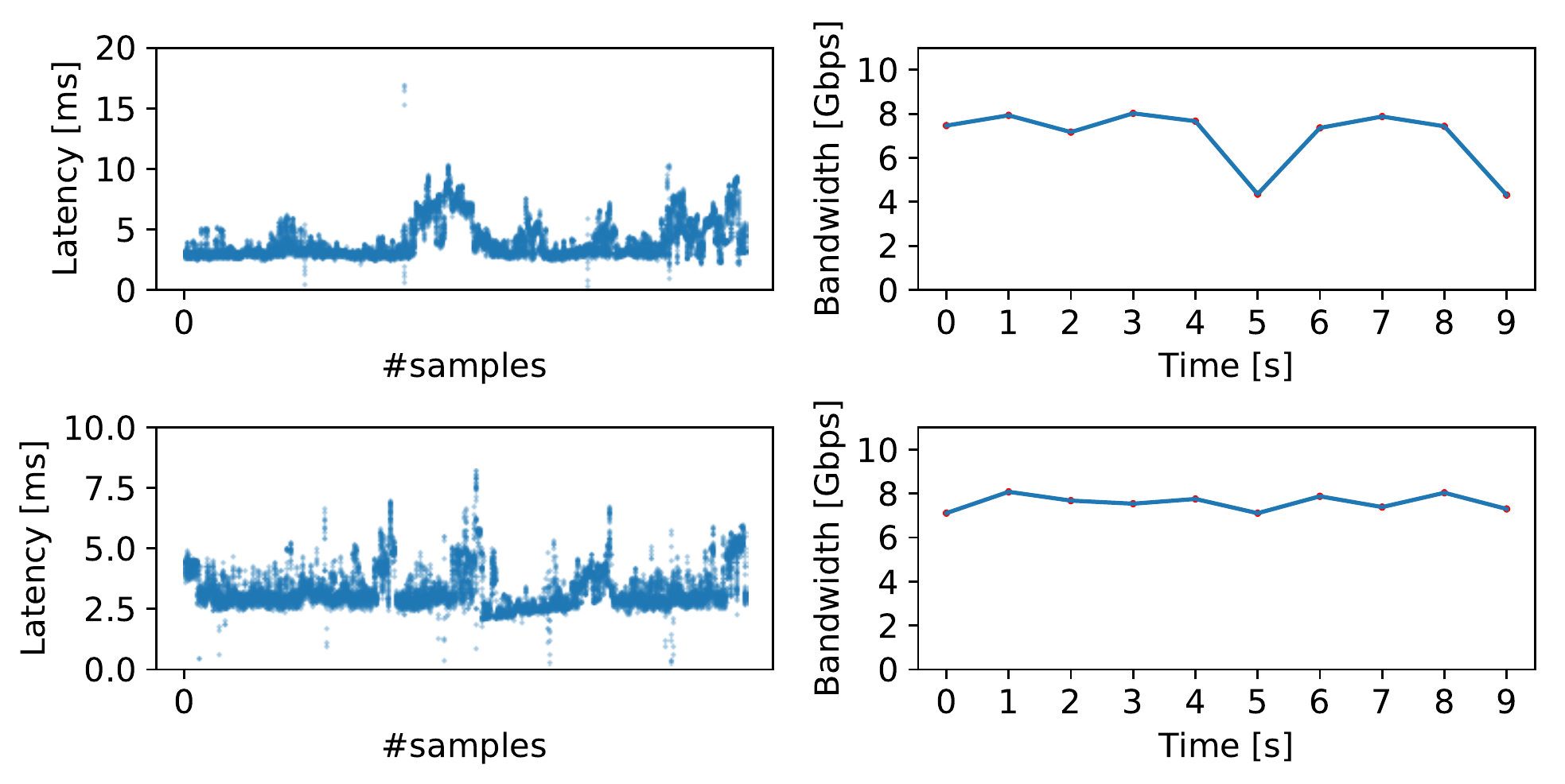}
    \vspace*{-0.2cm}
    \caption{Example of observed \google{} latency for a 10-second TCP sample on a \emph{4-core} instance. Left: RTT latency for TCP packets. Right: achieved \emph{iperf} bandwidth.}
    \label{fig:gce_latency}
    \vspace*{-0.3cm}
\end{figure}

\begin{figure}[t]
\centering
\subfigure {
		\includegraphics[width=0.41\linewidth]{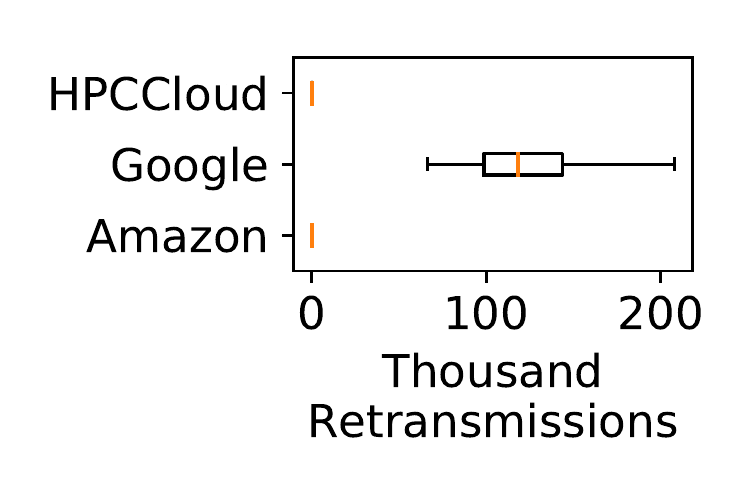}	
}
\subfigure{
		\includegraphics[width=0.41\linewidth]{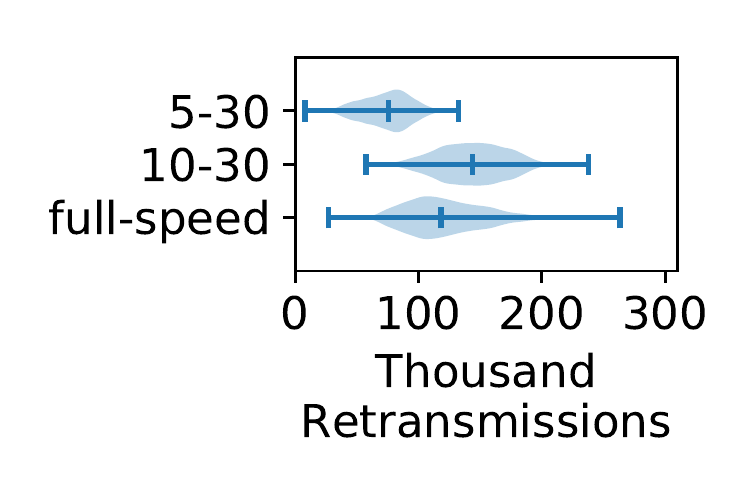}	
}
\vspace*{-0.4cm}
\caption{TCP retransmission analysis, summarized for all experiments presented before, in all clouds. Left: plots retransmissions as IQR boxplots, with the whiskers representing 1st and 99th percentiles; Right: violin plot for retransmissions in \google{}; thickness of the plot is proportional to the probability density of the data.}
\label{fig:retransmission_analysis}
\vspace*{-0.4cm}
\end{figure}

\begin{figure}
\centering
\subfigure[\amazon{}. \label{fig:aws_traffic}] {
		\includegraphics[width=0.47\columnwidth]{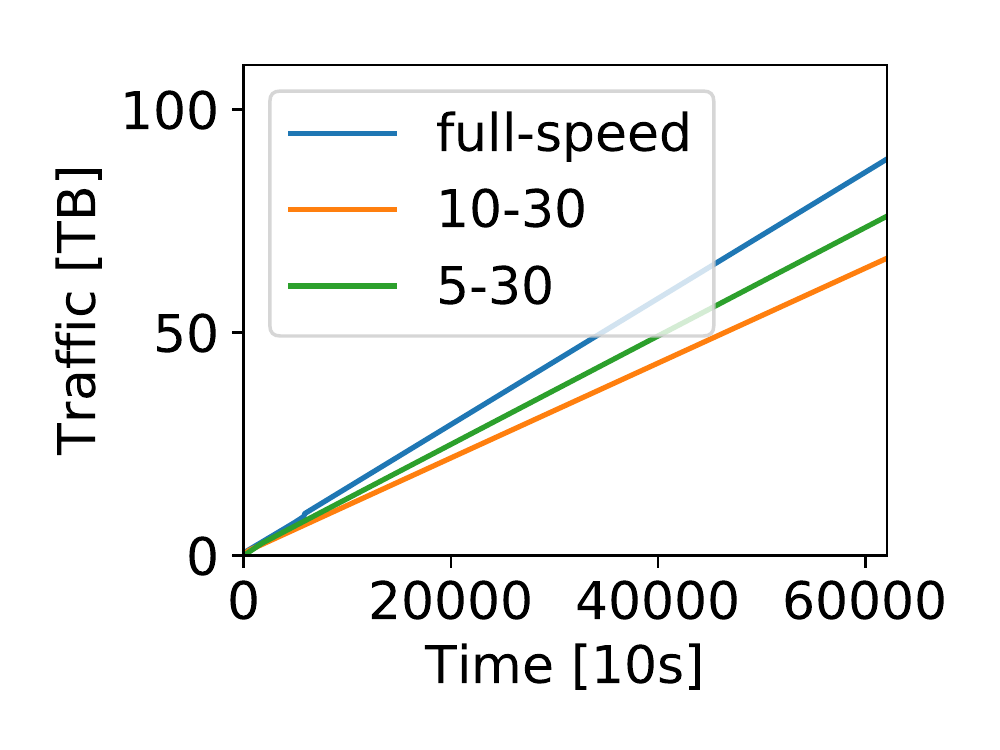}
}
\subfigure[\google{}. \label{fig:_google_traffic}]{
		\includegraphics[width=0.47\columnwidth]{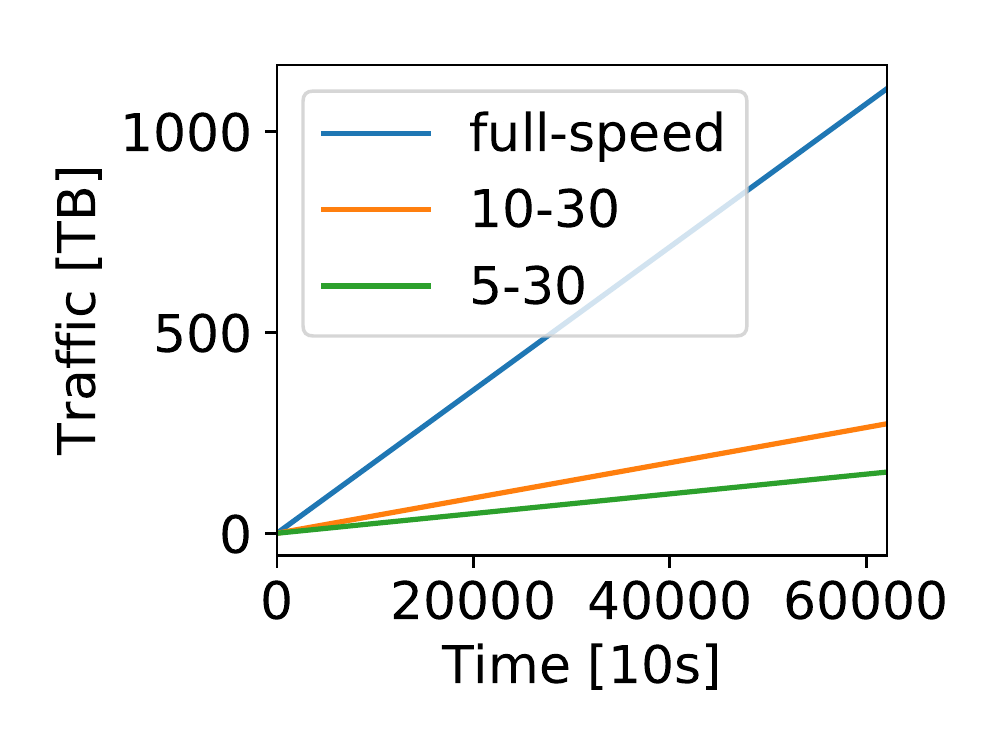}
}
\vspace*{-0.2cm}
\caption{The total amount of data transferred between the pairs of virtual machines involved in the three types of experiments performed. The total time is a week, while each point on the horizontal axis represents 10 seconds.}
\label{fig:traffic_analysis}
\vspace*{-0.4cm}
\end{figure}

\textbf{How rapidly does bandwidth vary?} 
Our analysis reveals that the level of \emph{measurement-to-measurement} variability is significant:
bandwidth in HPCCloud (\emph{full-speed}) and \google{} (\emph{5-30}) varies between consecutive 10-second measurements up to 33\% and 114\%, respectively. 
While a small sample may 
exhibit only modest fluctuations, the long-tailed distributions we 
observed here strongly suggest using the analysis techniques we discuss in Section~\ref{subsec:exp_stochastic_noise}.
\amazon{}'s variability is more particular
and policy-dependent, as we detail in Section~\ref{subsec:mec_policies}.

\subsection{Latency}
Commercial clouds implement their virtual networks using very different mechanisms and policies. We can see this in more detail by looking at the round-trip lantencies seen in \google{} and \amazon{}. 
We measure the application-observed TCP RTT, as this is what impacts the high-level networking stacks of big data frameworks. For our experiments, we run 10-second streams of \emph{iperf} tests, capturing all packet headers with \emph{tcpdump}. We perform an offline analysis of the packet dumps using \emph{wireshark}, which compares the time between when a TCP segment is sent to the (virtual) device and when it is acknowledged.
Our data was collected between August and September 2019. In total, we have over 50 million RTT datapoints. 

The behavior we observe is inherently different: \google{} exhibits latency in the order of milliseconds, with an upper limit of 10ms. \amazon{} generally exhibits faster sub-millisecond latency under typical conditions, but when the traffic shaping mechanism (detailed in Section~\ref{subsec:mec_policies}) takes effect, the latency increases by two orders of magnitude, suggesting large queues in the virtual device driver. Figure~\ref{fig:aws_latency} shows representative patterns of latency in the \amazon{} cloud, while Figure~\ref{fig:gce_latency} is representative of \google{}. Both figures plot latency as RTT packet data obtained from a 10-second TCP stream obtained running an \emph{iperf} benchmark. 

The behavior observed in the top half of Figure~\ref{fig:aws_latency} lasts for approximately ten minutes of full-speed transfer on \emph{c5.xlarge} instances. After this time, the VMs' bandwidth gets throttled down to about 1\,Gbps (bottom half of Figure~\ref{fig:aws_latency}), which also significantly increases latency. On \google{}, there is no throttling effect, but the bandwidth and latency vary more from sample to sample.

\subsection{Identifying Mechanisms and Policies}\label{subsec:mec_policies}
The behavior exhibited by the two commercial providers is notably different. We uncover mechanisms and policies for enforcing client QoS by performing extra analysis, depicted in Figures~\ref{fig:retransmission_analysis} and~\ref{fig:traffic_analysis}. The former plots the number of retransmissions per experiment 
(part (a)) and a zoomed-in view of \google{}
(part (b)). 
While both \amazon{} and \surf{} have a negligible number of retransmissions, retransmission are common in \google{}: roughly 2\% per experiment. 

Figure~\ref{fig:traffic_analysis} plots the total amount of traffic for \amazon{} 
and \google{} 
over the entire duration of our experiments. It is clear that in \google{}'s case the amount of traffic generated by \emph{full-speed} is orders of magnitude larger than for the intermittent access patterns. In \amazon{}'s case, the total amount of data sent for all three kinds of experiments is roughly equal. By corroborating this finding the more fine-grained experiments we performed presented in Figure~\ref{fig:aws_latency}, and other empirical studies~\cite{persico2015measuring,wang2017using}, we find that this provider uses a \emph{token-bucket} algorithm to allocate bandwidth to its users. 

\textbf{Token-Bucket Analysis.} The token-bucket algorithm operation can be explained as follows. When a VM is provided to the user, its associated \emph{bucket} holds a certain amount of tokens (i.e., a budget). This budget is allowed to be spent at a high rate (i.e., 10\,Gbps). When the budget is depleted (e.g., after about 10 minutes of continuous transfer on a \emph{c5.xlarge} instance, the QoS is limited to a low rate (e.g., 1\,Gbps). The bucket is also subject to a replenishing rate that we empirically found to be approximately 1 Gbit token per second, i.e., every second users receive the amount of tokens needed to send 1Gbit of data at the high (10Gbps) rate. Once the token bucket empties, transmission at the capped rate is sufficient to keep it from filling back up. The user must \emph{rest} the network, and re-filling the bucket completely takes several minutes.

\begin{figure}[tp]
    \centering
    \includegraphics[width=0.9\columnwidth]{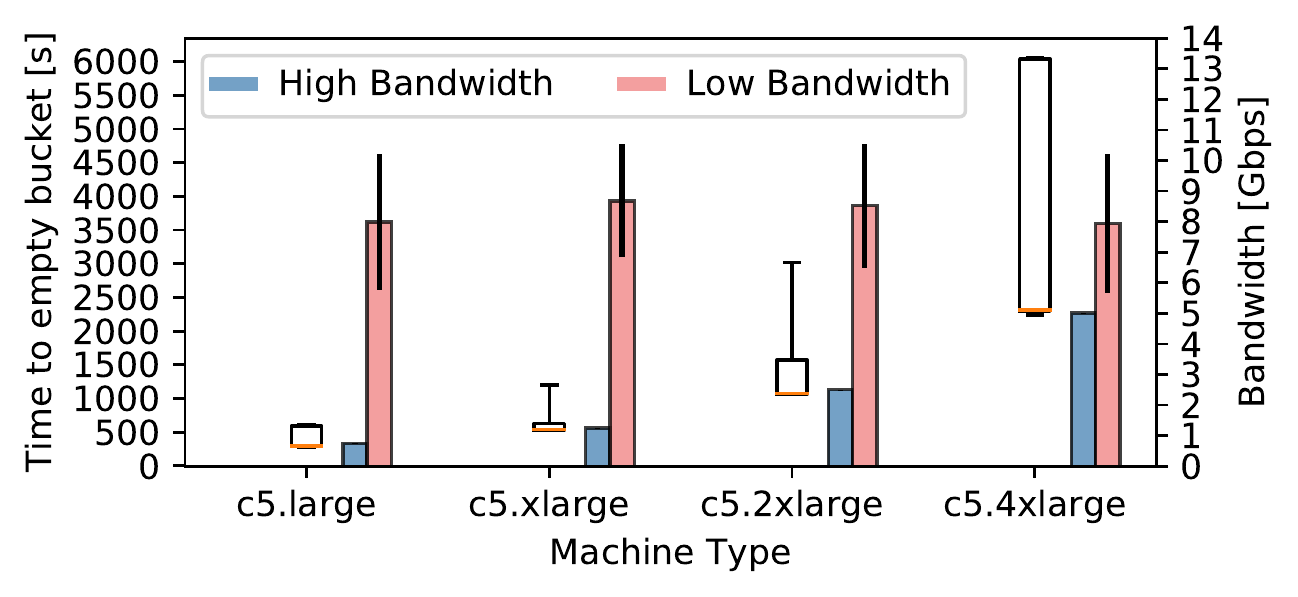}
    \vspace*{-0.2cm}
    \caption{The token-bucket parameters identified for several instances of \amazon{} \emph{c5.*} family. The elapsed time to empty the token bucket is depicted with boxplots associated with \emph{left} vertical axis. The \emph{high} and \emph{low} bandwidths of the token bucket are depicted with bar plots with whiskers and are associated with the \emph{right} vertical axis.}
    \label{fig:token_bucket_sizes}
    \vspace*{-0.2cm}
\end{figure}

We analyze the behavior of multiple types of VMs from the \emph{c5.*} family, and find that their token-bucket parameters differ. More expensive machines benefit from larger initial budgets, as well as higher bandwidths when their budget depletes. Figure~\ref{fig:token_bucket_sizes} plots the token-bucket parameter analysis for four VMs of the \emph{c5.*} family. For each VM type, we ran an \emph{iperf} test continuously until the achieved bandwidth dropped significantly and stabilized at a lower value. For each instance type, we ran 15 tests. Figure~\ref{fig:token_bucket_sizes} shows the time taken to empty the token bucket, the \emph{high} (non-empty bucket) bandwidth value, and the \emph{low} (empty bucket) bandwidth value. As the \emph{size} (i.e., number of cores, amount of memory etc,) of the VM increases, we notice that the \emph{bucket size} and the \emph{low} bandwidth increase proportionally. However, as the magnitude of the boxplots suggests, as well as the error bars we plotted for the \emph{high} bandwidth, these parameters are not always consistent for multiple incarnations of the same instance type. 

\begin{figure}[tp]
    \centering
    \includegraphics[width=0.9 \columnwidth]{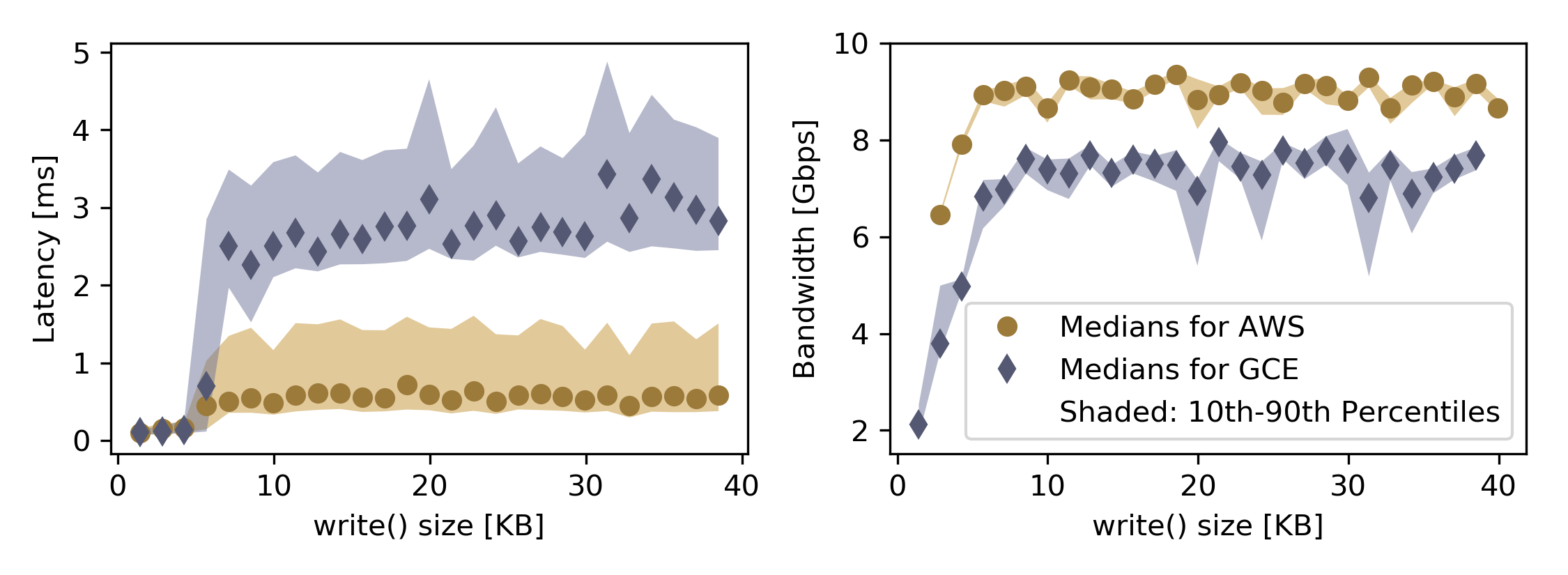}
    \vspace*{-0.2cm}
    \caption{Measured latency and bandwidth for Amazon EC2 (\emph{c5.xlarge}) and GCE (4-core VM with advertised $8Gbps$) instances
    as functions of the  \textmd{\texttt{write()}} size.}
    \label{fig:bw_lat}
    \vspace*{-0.2cm}
\end{figure}

\textbf{Virtual NIC Implementations.} We found that differences in EC2 and GCE's implementations of virtual NICs can lead to significantly different observed behavior. EC2's virtual NICs advertise an MTU of 9000 bytes, a standard “jumbo frame” size. GCE's only advertise an MTU of 1500 bytes (standard Ethernet frame size), but instead enable TCP Segmentation Offloading (TSO), in which the NIC accepts larger “packets” from the device driver, but then breaks them down into smaller Ethernet frames before transmission (we do not know whether this occurs at the virtual or physical NIC in GCE's implementation). Both of these techniques serve the same basic function—reducing overhead by sending fewer, larger packets on the virtual NIC, but result in different observable behavior on the host, and the details of this behavior depend heavily on the application and workload.

The most striking effect is the way that the size of the \texttt{write()}s done by the application affects latency and packet retransmission. Figure~\ref{fig:bw_lat} plots the effects of the \texttt{write()} size on latency and bandwidth. On EC2, the size of a single “packet” tops out at the MTU of 9K, whereas on GCE, TSO can result in single “packet” at the virtual NIC being as large as 64K in our experiments. With such large “packets,” perceived latency increases greatly due to the higher perceived “transmission time” for these large packets. The number of transmissions also goes up greatly, presumably due to limited buffer space in the bottom half of the virtual NIC driver or tighter bursts on the physical NIC. In practice, the size of the “packets” passed to the virtual NIC in Linux tends to equal to the \texttt{write} on the socket (up to the cap noted above). This makes the observed behavior (and thus repeatability, and the ability to generalize results between clouds) highly application dependent. It is also worth noting that all streams are affected when one stream sends large “packets”, since they share a queue in the virtual device driver. On GCE, when we limited our benchmarks to \texttt{write}s of 9K, we got near-zero packet retransmission and an average RTT of about 2.3$ms$. When the benchmark used its default \texttt{write()} size of 128K, we saw the hundreds of thousands of retransmission shown in Figure~\ref{fig:retransmission_analysis} and latencies as high as 10$ms$.

\section{Performance Reproducibility For Big Data Applications}\label{sec:mechanisms_effect}\label{sec:postgigabit_bigdata}

Having looked at low-level variability in bandwidth and latency, we now move “up” a level to applications and workloads.
Our main findings are:
\begin{description}[noitemsep,nosep,leftmargin=-0.03cm]
\item[F4.1] Under variability resembling \google{} and \surf{}, which can be modeled as stochastic noise, reproducible experiments can be obtained using sufficient repetitions and sound statistical analyses. 
\item[F4.2] Application transfer patterns exhibit non-trivial interactions with token-bucket network traffic shapers. Depending on the bucket budget and the application, significant application performance variability is incurred.
\item[F4.3] Token-bucket traffic shapers in conjunction with (imbalanced) big data applications can create stragglers.
\item[F4.4] In long-running cloud deployments that have incurred large amounts of varied network traffic, it is highly difficult to predict application performance, as it is dependent on the state of the individual nodes' remaining token-bucket budgets.

\end{description}

\subsection{Experiments and Stochastic Noise}\label{subsec:exp_stochastic_noise}

\begin{figure}[t]
    \centering
    \begin{subfigure}{\small{(a) Median Performance for K-Means in Google Cloud.}}
       \includegraphics[width=0.9\columnwidth]{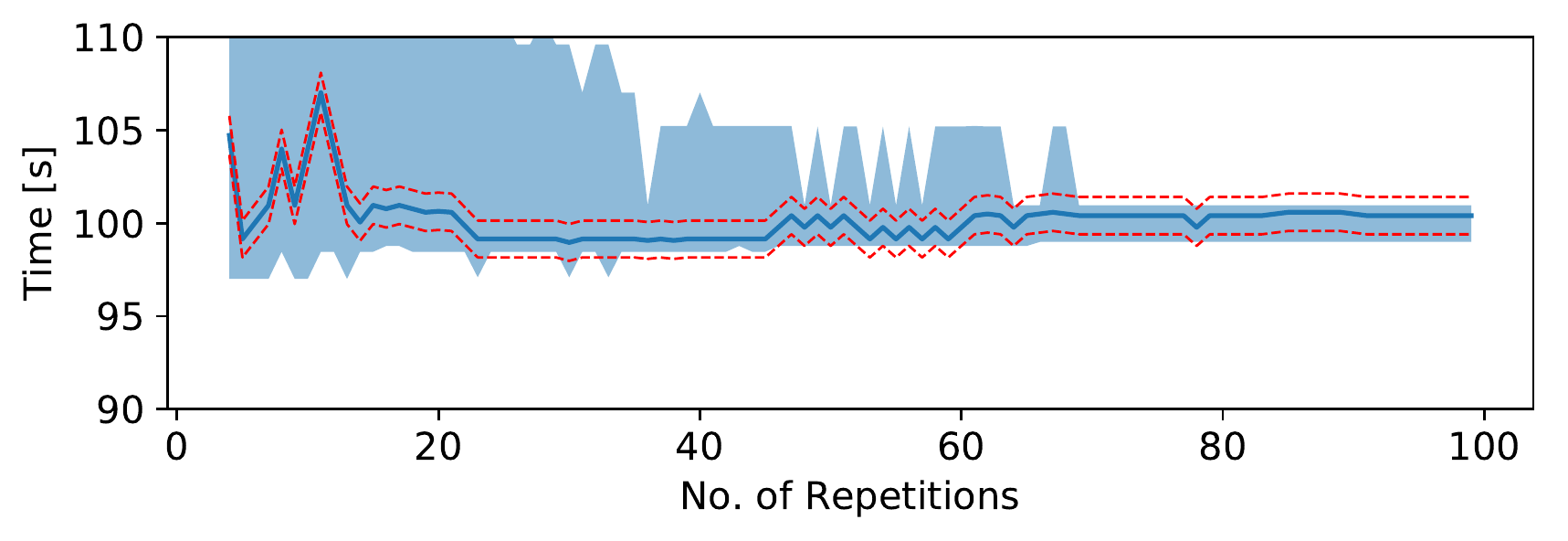}
           \vspace{-0.2cm}
       \label{fig:google-kmeans} 
    \end{subfigure}
    \begin{subfigure}{\small{(b) Median Performance for TPC-DS Q65 in \surf{}.}}
       \includegraphics[width=0.9\columnwidth]{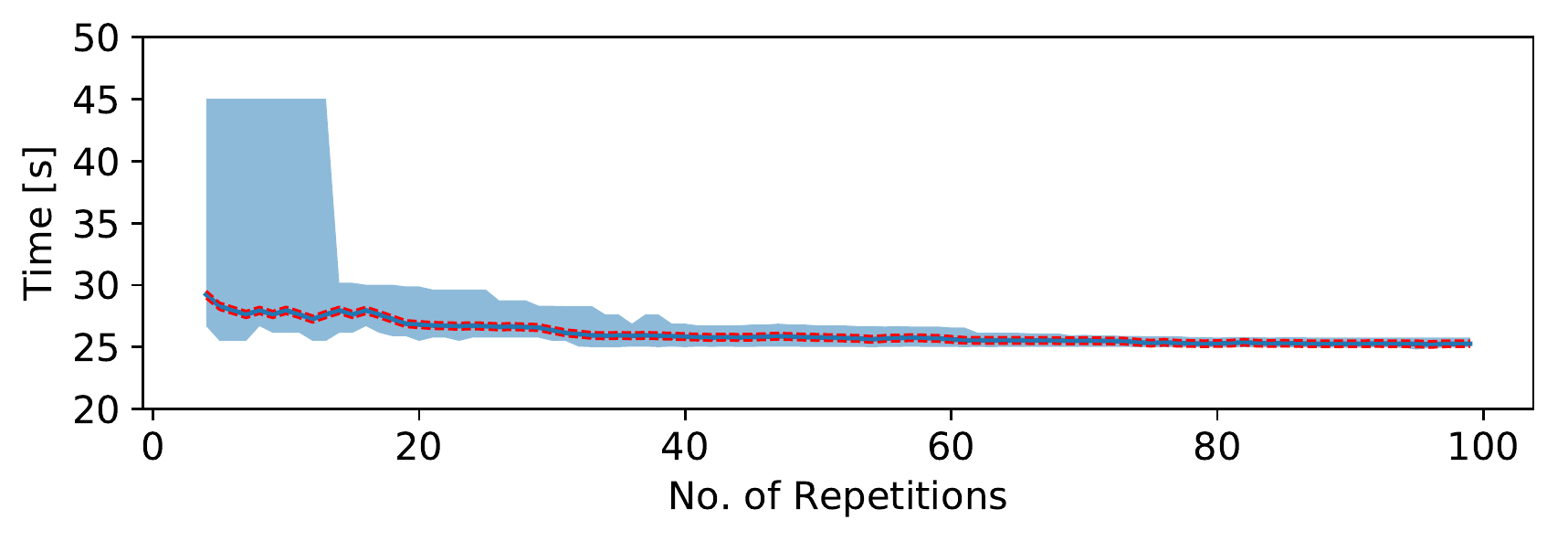}
       \label{fig:surf-tpcds}
    \end{subfigure}
    \vspace*{-0.4cm}
    \caption{CONFIRM analysis for K-Means and TPC-DS Q65 on \google{} and \surf{}. Median estimates (blue thick curve), 95\% nonparametric confidence intervals (light blue filled space), and 1\% error bounds (red dotted curves). Vertical axis not starting at 0 for visibility.}
    \label{fig:stochastic_variability}
    \vspace*{-0.1cm}
\end{figure}

As detailed in Section~\ref{sec:netw_var}, the behavior of the network performance variability for \google{} and \surf{} is closer in nature to stochastic variability given by transient conditions in the underlying resources, such as noisy neighbors. To achieve reproducible experiments under such conditions, system designers and experimenters need to carefully craft and plan their tests, using multiple repetitions, and must perform sound statistical analyses.

We ran several HiBench~\cite{huang2010hibench} and TPC-DS~\cite{nambiar2006making} benchmarks directly on the \google{} and \surf{} clouds and report how many repetitions an experimenter needs to perform in order to achieve trustworthy experiments. While it is true that running experiments directly on these clouds we cannot differentiate the effects of network variability from other sources of variability, the main take-away message of this type of experiment is that this kind of stochastic variability can be accounted for with proper experimentation techniques.

On the performance data we obtained, we performed a CONFIRM~\cite{maricq2018taming} analysis, which helps to predict how many repetitions an experiment will require to achieve a desired confidence interval. Figure~\ref{fig:stochastic_variability} presents our findings, showing that for these two common benchmarks, it can take 70 repetitions or more to achieve 95\% confidence intervals within 1\% of the measured median. As we saw in Section~\ref{sec:litsurv}, this is far more repetitions than are commonly found in the literature: most papers are on the extreme left side of this figure, where the confidence intervals are quite wide. This points to the need for stronger experiment design and analysis in our community.

\begin{figure}
    \centering
    \includegraphics[width=0.9\columnwidth]{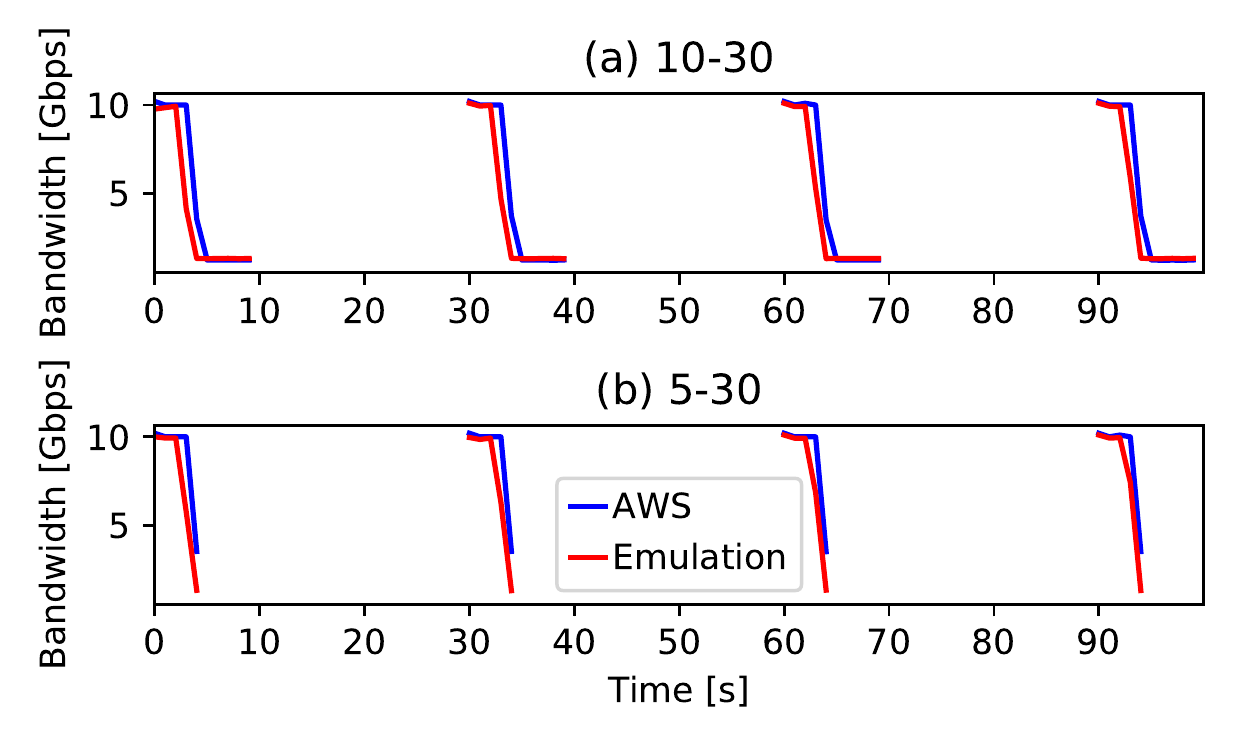}
    \vspace*{-0.4cm}
    \caption{Validation of the emulation of the token-bucket policy of \amazon{}. The similar aspect of the two curves indicates that emulation is high-quality.}
    \label{fig:aws_emulation_validation}
    \vspace*{-0.2cm}
\end{figure}

\begin{table}
\caption{Big data experiments on modern cloud networks.}
\resizebox{0.98\columnwidth}{!}{
\begin{tabular}{@{}ccccc@{}}
\toprule
\textbf{Workload} & \textbf{Size} & \textbf{Network} & \textbf{Software} & \textbf{\#Nodes} \\ \midrule
HiBench~\cite{huang2010hibench}           & BigData      & \begin{tabular}[c]{@{}c@{}}Token-bucket,\\ Figure~\ref{fig:aws_emulation_validation}\end{tabular}                                                          & \begin{tabular}[c]{@{}c@{}}Spark 2.4.0, \\ Hadoop 2.7.3\end{tabular}            & 12                  \\
TPC-DS~\cite{nambiar2006making}            & SF-2000        & \begin{tabular}[c]{@{}c@{}}Token-bucket,\\ Figure~\ref{fig:aws_emulation_validation}\end{tabular}                                                           & \begin{tabular}[c]{@{}c@{}}Spark 2.4.0,\\ Hadoop 2.7.3\end{tabular}             & 12                  \\ \bottomrule
\end{tabular}
}
\label{tab:exp_postgigabit_era}
\vspace*{-0.25cm}
\end{table}

\begin{figure}[t]
\centering
\includegraphics[width=0.99\columnwidth]{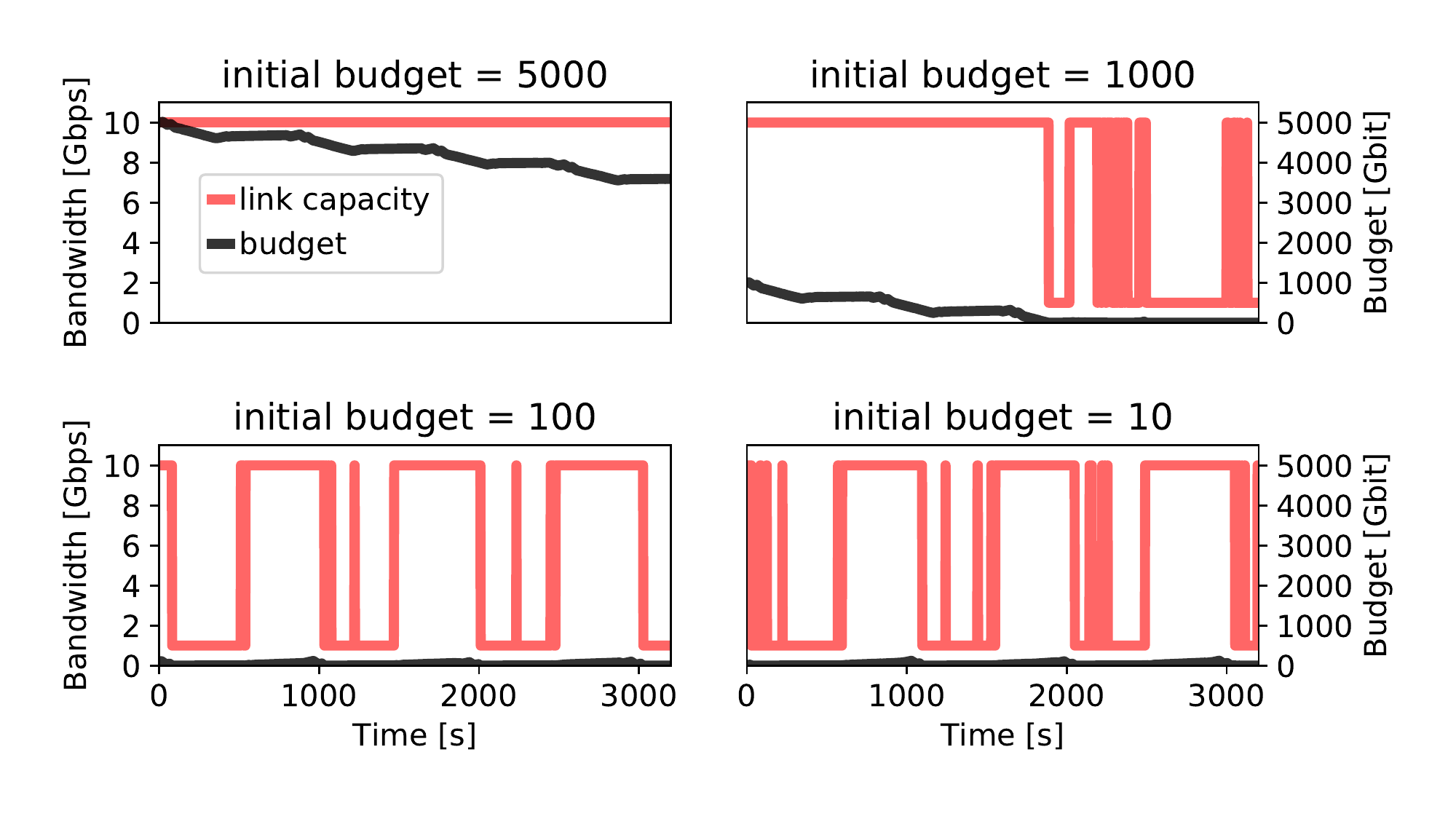}
\vspace*{-0.4cm}
\caption{Link capacity allocated when running Terasort on a token bucket. Left vertical axis shows the link capacity; right vertical axis shows the token bucket budget. Budget depletes due to application network transfers.}
\label{fig:terasort_traffic_analysis}
\vspace*{-0.1cm}
\end{figure}

\begin{figure}[t]
    \centering
    \includegraphics[width=0.97\columnwidth]{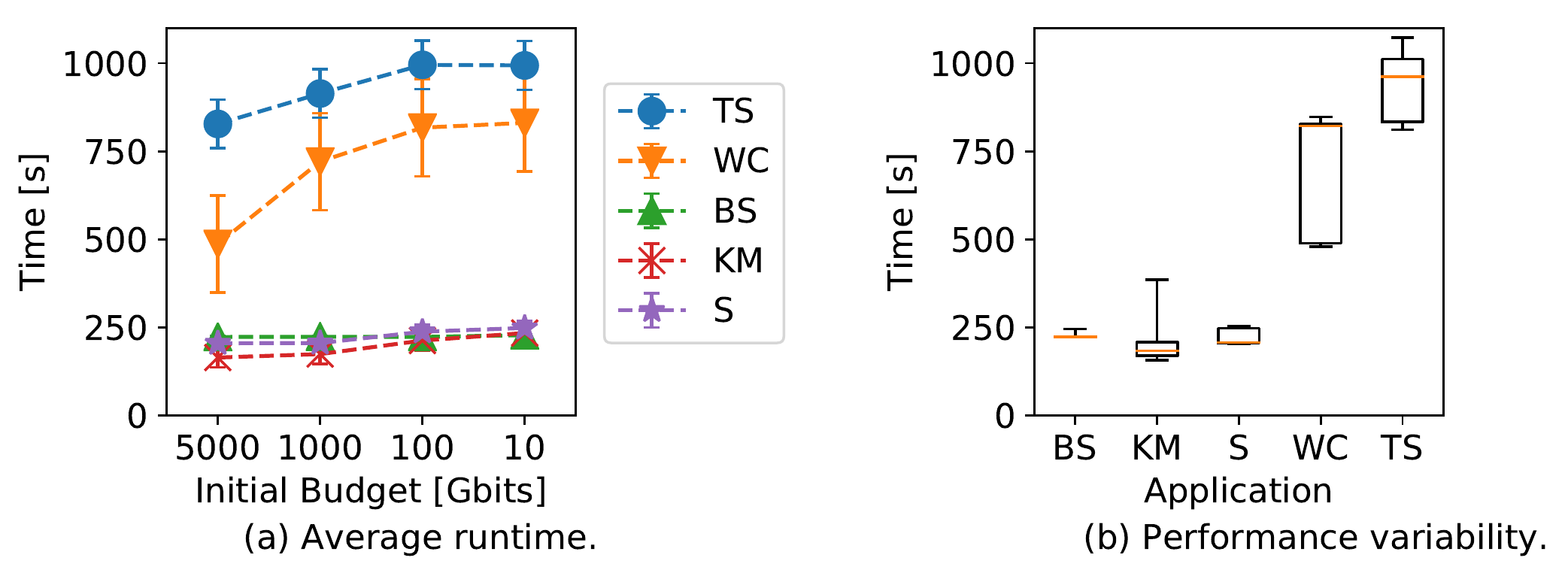}
    \vspace*{-0.2cm}
    \caption{HiBench average runtime (left) and performance variability (right), plotted as IQR box (whiskers represent 1st and 99th percentiles), induced by token bucket budget variability. The more network-dependent applications are affected more by lower budgets.}
    \label{fig:hibench_aws_slowdown}
    \vspace*{-0.2cm}
\end{figure}

\begin{figure*}[t]
    \centering
    \includegraphics[width=0.92\textwidth]{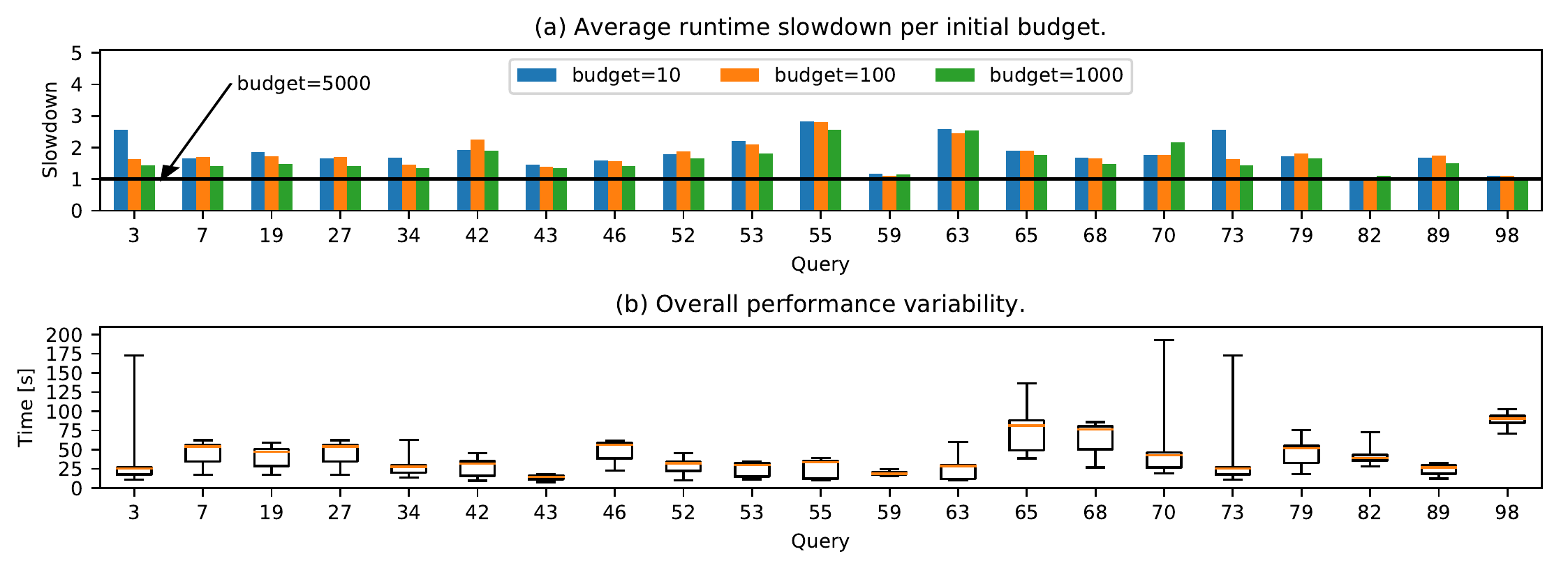}
    \vspace*{-0.2cm}
    \caption{TPC-DS average runtime slowdown per query depending on initial budget (top); overall performance variability, summarized over initial budgets (bottom), plotted as IQR box; whiskers represent 1st and 99th percentiles.}
    \label{fig:tpcds_aws_slowdown}
    \vspace*{-0.1cm}
\end{figure*}

\subsection{Experiments and Token-Buckets}
In contrast to \google{} and \surf{}, the  \emph{token-bucket} shaping policy of \amazon{} is \emph{not} stochastic noise, and needs in-depth analysis. Because token-bucket behavior is dependent on past network access patterns, \textbf{an application influences not only its own runtime, but also future applications' runtimes}.

\textbf{Token-bucket Emulator.} We decided to emulate the behavior of \amazon{} token-bucket instead of directly running applications in this cloud. We believe this type of experimentation is superior to the other two alternatives: (i) simulation, or (ii) directly running applications on the cloud. For the former, we believe the behavior of big data applications under network performance variability is far too subtle and complex to properly simulate while modeling and capturing all possible variables.
For the latter, we perform the emulation in an isolated setup, i.e., a private cluster, that does not share resources. This allows us to test in isolation the effects of network performance variability, excluding as much as possible all other sources of variability one could encounter in a cloud (e.g., CPU, memory bandwidth, I/O etc.). If we were to directly run applications in a cloud, it would have been difficult to separate the effects of network variability from, for example, the effects of CPU variability. 

We built a network emulator based on the Linux \texttt{tc}~\cite{hubert2002linux} facility. Figure~\ref{fig:aws_emulation_validation} plots the real-world behavior encountered in \amazon{} in comparison with our emulation. This experiment is a zoomed-in view of the experiment in Section~\ref{subsec:var_postgbit_networks}, where our servers were communicating for either five or ten seconds, then slept for 30 seconds. At the beginning of each experiment, we made sure that the token-bucket budget is nearly empty. During the first few seconds of the experiment the token-bucket budget gets completely exhausted. For each sending phase of 5 or 10 seconds, the system starts at a high QoS (10\,Gbps bandwidth), after a few seconds the budget is emptied, and the system drops to a low QoS (1\,Gbps).

\textbf{Experiment Setup.} We perform the experiments described in Table~\ref{tab:exp_postgigabit_era} on a 12-node cluster. Each node has 16 cores, 64GB memory, a 256GB SSD, and FDR InfiniBand network. Using the emulator presented in Figure~\ref{fig:aws_emulation_validation}, we run on the emulated \amazon{} token-bucket policy all applications and queries in the HiBench~\cite{huang2010hibench} and TPC-DS~\cite{nambiar2006making} benchmark suites. The emulated setup is that of the \emph{c5.xlarge} instance type, which typically sees a high bandwidth of 10\,Gbps, and a low bandwidth of 1\,Gbps. Throughout our experiments we vary the token bucket budget to assess its impact on big data applications. We run each workload a minimum of 10 times for each token-bucket configuration and report full statistical distributions of our experiments.


\textbf{Token-bucket-induced Performance Variability.} One important parameter for the token-bucket is its budget: the number of tokens available at a certain moment in time. This is highly dependent on the previous state of the virtual machine (i.e., how much network traffic has it sent recently), and has a large impact on the performance of future deployed applications. Note that it is difficult to estimate the currently-available budget for anything other than a “fresh” set of VMs: each VM has its own token bucket, the remaining budget is a function of previous runs, and, as we saw in Figure~\ref{fig:token_bucket_sizes} the constants controlling the bucket are not always identical.
%

Application performance is highly dependent on the budget, and deployments with smaller budgets create more network performance variability. Figure~\ref{fig:terasort_traffic_analysis} shows the network traffic behavior of the Terasort application with different initial budgets. For each budget, the subfigures show the application network profile for 5 consecutive runs. We notice a strong correlation between small budgets and network performance variability: there is much more variability for budgets~$\in \{10, 100\}$ Gbits, than for budgets~$\in \{1000, 5000\}$ Gbits.

Figure~\ref{fig:hibench_aws_slowdown} shows how this effect manifests in the runtimes of HiBench: it plots the average application runtime (left) over 10 runs for budgets $\in \{10, 100, 1000, 5000\}$ Gbits, and the performance variability over the same budgets (right). For the more network-intensive applications (i.e., TS, WC), the  initial state of the budget can have a 25\%--50\% impact on performance.

A similar behavior is observed for the TPC-DS benchmark suite. Figure~\ref{fig:tpcds_aws_slowdown} shows the query sensitivity to the token budget and the variability induced by different budget levels. Figure~\ref{fig:tpcds_aws_slowdown}(a) plots average runtime \addition{slowdown} for 10-run sets of TPC-DS queries for budgets $\in \{10, 100, 1000\}$ Gbits, \addition{compared to the 5000Gbit budget}. For all queries, larger budgets lead to better  performance. Figure~\ref{fig:tpcds_aws_slowdown}(b) plots the performance variability over all tested budgets. Queries with higher network demands exhibit more sensitivity to the budget and hence higher performance variability.

These results clearly show that if the system is left in an unknown state (e.g., a partially-full token bucket, left over from previous experiments), the result is likely to be an inaccurate performance estimate. 
Evidence from Figures~\ref{fig:hibench_aws_slowdown}(b) and \ref{fig:tpcds_aws_slowdown}(b) strongly supports this, as performance varies widely for the network-intensive queries and applications depending on the token-bucket budget.

\begin{figure}[t]
    \centering
    \includegraphics[width=0.99\columnwidth]{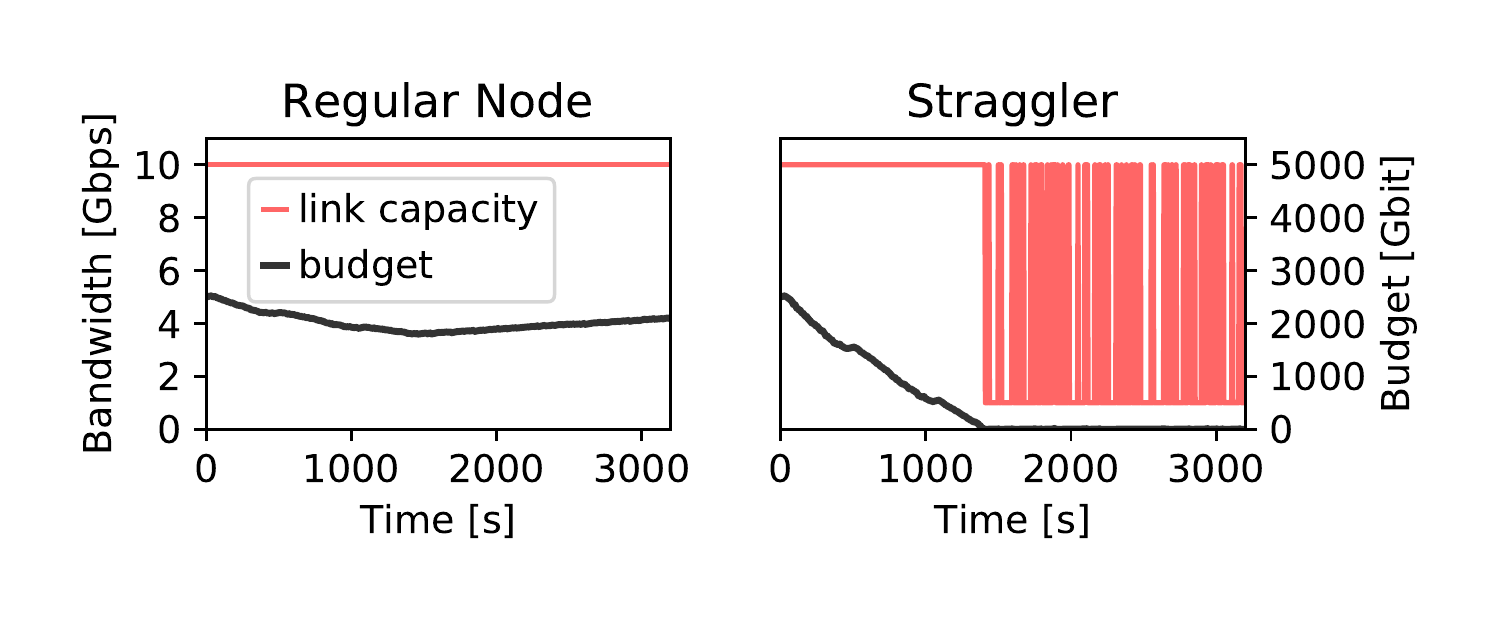}
    \vspace*{-0.64cm}
    \caption{Link capacity allocation for TPC-DS on a token-bucket network, with initial budget = 2500Gbit. Regular node network utilization (left); straggler node (right).}
    \label{fig:straggler_tpcds_aws}
    \vspace*{-0.45cm}
\end{figure}

\textbf{Token-bucket-induced Stragglers.} 
Non-trivial combinations of token-bucket budgets, application scheduling imbalances, and network access patterns lead to straggler nodes. Figure~\ref{fig:straggler_tpcds_aws} shows that for budget = $2500$\,Gbits and application TPC-DS, the application gets slowed down by a straggler: all nodes but one in the deployment do not deplete their budgets completely, thus remaining at a high bandwidth QoS of 10Gbps. However, there is one node on which the token-bucket budget is depleted, causing its bandwidth to get limited to 1Gbps. Exacerbating the variability, the behavior is not even fixed: this node oscillates between high and low bandwidths in short periods of time. Such unpredictable behavior leads to both performance variability of the entire setup and also poor experiment reproducibility.

\textbf{Repeatable experiments and token-buckets.} Token-bucket policies for enforcing network QoS can have unexpected and detrimental impacts
on sound cloud-based experimentation. To explore this, we compute medians and their nonparametric confidence intervals (CIs), similar to the work by Maricq et al.~\cite{maricq2018taming}, across a number of initial token budgets. Figure~\ref{fig:token_bucket_CI} plots median estimates for two TPC-DS queries, along with 95\% CIs and 10\% error bounds around medians. Repetitions of the experiments are independent: each one runs on fresh machines with flushed caches, and at the the beginning of each repetition, we reset the token budget. 
We reduce this initial budget over time to emulate the effects that previous experiments can have on subsequent ones: what this models is an environment in which many different experiments (or repetitions of the same experiment) are run in quick succession. This is likely to happen when running many experiments back-to-back in the same VM instances.

Query 82 (in the top of Figure~\ref{fig:token_bucket_CI}) is agnostic to the token budget. Running  more repetitions of this experiment tightens the confidence intervals, as is expected in CI analysis. In contrast, query 65 (in the bottom of the figure) depends heavily on the bucket budget; as a result, as we run more experiments, depleting the bucket budget, the query slows down significantly, and the initial CI estimates turn out to be inaccurate. In fact, the CIs \emph{widen} with more repetitions, which is unexpected for this type of analysis. This is because the token bucket breaks the assumption that experiments are independent: in this model, more repetitions deplete the bucket that the next experiment begins with. These two queries represent extremes, but, as shown in the bar graph at the bottom of the figure, 80\% of all queries we ran from TPC-DS suffer effects like Query~65: most produce median estimates that are more than 10\% incorrect by the time we fully deplete the budget.


This demonstrates that, when designing experiments, we cannot simply rely on the intuition that more repetitions lead to more accurate results: we must ensure that factors hidden in the cloud infrastructure are reset to known conditions so that each run is truly independent. Others have shown that cloud providers use token buckets for other resources such as CPU scheduling~\cite{wang2017using}. 
This affects cloud-based experimentation, as the state of these token buckets is not directly visible to users, nor are their budgets or refill policies. 
%
%
%

\section{Summary: Is Big Data Performance Reproducible in Modern Cloud Networks?}\label{sec:protocol}

%

We return to our two basic questions: (1) \emph{How reproducible are big data experiments in the cloud?}; and (2) \emph{What can experimenters do to make make sure their experiments are meaningful and robust?} Our findings are:

\textbf{F5.1: Network-heavy experiments run on different clouds cannot be directly compared.}

Building a cloud involves trade-offs and implementation decisions, especially at the virtualization layer.
Some of these decisions are well-documented by the platforms~\cite{aws_enhanced_networking,google_networking}, but others, including the ones we have examined in this paper, are not. Unfortunately, these differences can cause behaviors that result in different application performance, such as the bandwidth differences seen in Figure~\ref{fig:traffic_analysis} or the latency effects seen in Figure~\ref{fig:bw_lat}. 

Both of these effects are rather large, and are dependent on factors such as the size of the applications' write buffer and specific patterns of communication. While these decisions presumably serve the clouds' commercial customers well, they complicate things for those who are trying to draw scientific conclusions; when comparing to previously-published performance numbers, it is important to use the same cloud to ensure that differences measured are that of the systems under test, and not artifacts of the cloud platform.
Running on multiple clouds, can, however, be a good way to perform sensitivity analysis~\cite{jain1990art}: by running the same system with the same input data and same parameters on multiple clouds, experimenters can reveal how sensitive the results are to the choices made by each provider.

\begin{figure}[t]
    \centering
    \includegraphics[width=0.99\columnwidth]{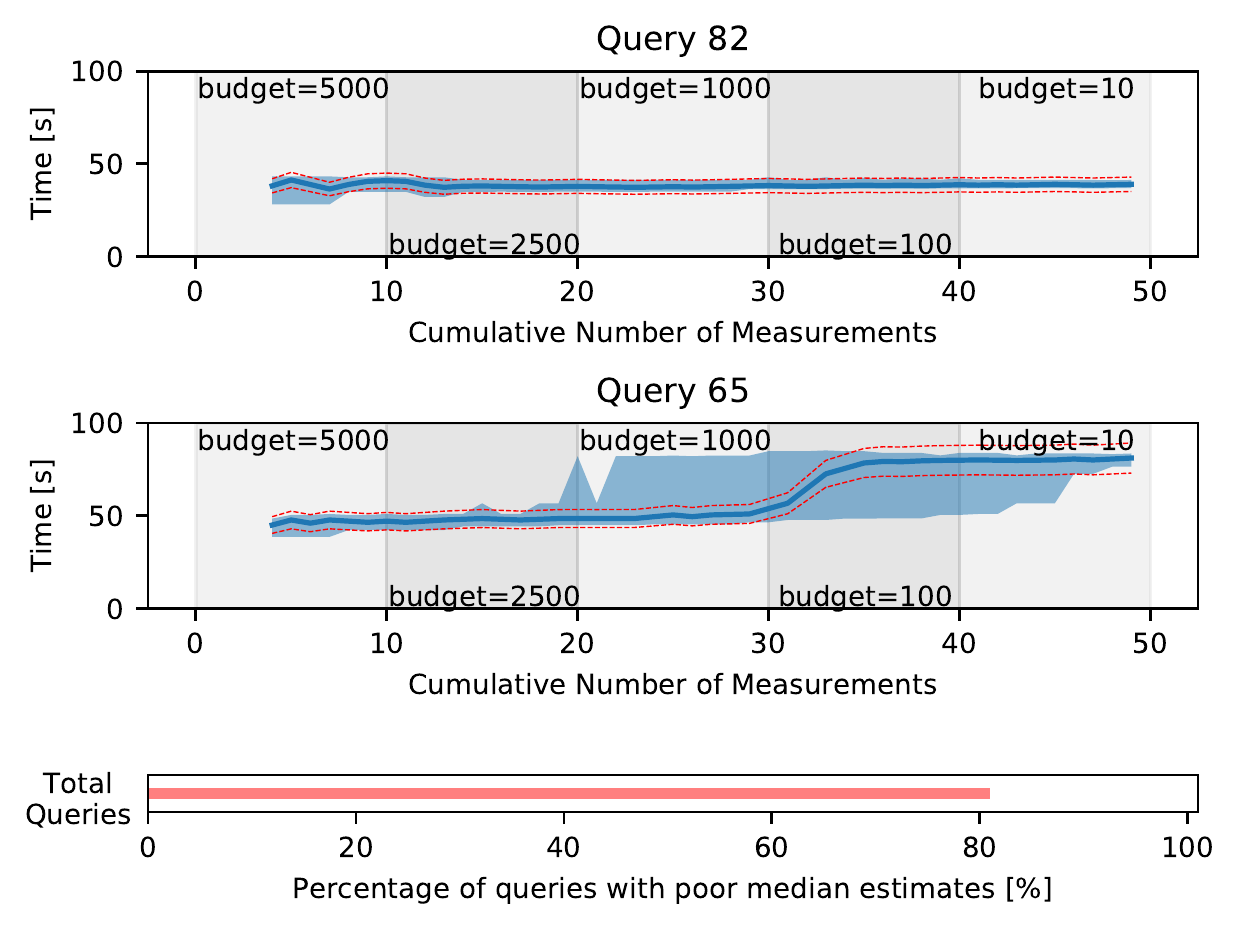}
    \vspace*{-0.2cm}
    \caption{Median estimates (blue thick curve), 95\% nonparametric confidence intervals (light blue filled space), and 10\% error bounds (red dotted curves) for running two TPC-DS queries, over 5 token-bucket budgets. Bottom: number of queries for which we cannot achieve tight confidence intervals and accurate median estimates.}
    \label{fig:token_bucket_CI}
    \vspace*{-0.2cm}
\end{figure}

\textbf{F5.2: Even within a single cloud, it is important to establish baselines for expected network behavior. These baselines should be published along with results, and need to be verified before beginning new experiments.}

Because cloud providers' policies can be opaque, and implementation details can change over time, it is possible for changes to invalidate over time experiments within the same cloud. For example, after several months of running experiments in \amazon{}, we began encountering new behavior: prior to August 2019, all \texttt{c5.xlarge} instances we allocated were given virtual NICs that could transmit at 10\,Gbps. Starting in August, we started getting virtual NICs that were capped to 5\,Gbps, though not consistently (this behavior is part of the underlying cause of the distributions in Figure~\ref{fig:token_bucket_sizes}). The reasons for this are not clear, and we have no way to know whether the “new” behavior is a transient effect in response to increased congestion that month or a new, permanent policy.

If one can establish baseline expectations for how the platform will perform, and incorporate checks for them into the experimental process~\cite{jimenez2017popper}, one can at least detect when changes have occurred. Experimenters should check, through micro-benchmarks, whether specific cloud resources (e.g., CPU, network) are subject to provider QoS policies. 

As opposed to contention-related variability, this type of variability is deterministic under carefully selected micro-benchmarks. In the network, these microbenchmarks should at a minimum include base latency, base bandwidth, how latency changes with foreground traffic, and the parameters to bandwidth token-buckets, if they are present. Furthermore, when reporting experiments, always include these \emph{performance fingerprints} together with the actual data, as possible changes in results in the future could be explained by analyzing the micro-benchmark logs.

\textbf{F5.3: Some cloud network variability (in particular, interference from neighbors) can be modeled as stochastic noise, and classic techniques from statistics and experiment design are sufficient for producing robust results; however, this often takes more repetitions than are typically found in the literature.}

Standard statistical tools such as ANOVA and confidence intervals~\cite{jain1990art,Boudec2011Performance,maricq2018taming} are effective ways of achieving robust results in the face of random variations, such as those caused by transient “noisy neighbors”; however, in order to be effective, they require many repetitions of an experiment, and, as we saw in Section~\ref{sec:litsurv}, this bar is often not met in the literature. The more variance, the more repetitions are required, and as we saw in Figures~\ref{fig:perf_var_aws}, \ref{fig:perf_var_google}, and~\ref{fig:perf_var_surf}, network variance in the cloud can be rather high, even under ‘ideal’ conditions. An effective way to determine whether enough repetitions have been run is to calculate confidence intervals for the median and tail, and to test whether they fall within some acceptable error bound (e.g., 5\% of value they are measuring).

\textbf{F5.4: Other sources of variability cause behavior that breaks standard assumptions for statistical analysis, requiring more careful experiment design.}

Some of the variability we have seen (e.g., Figures~\ref{fig:bw_lat}, \ref{fig:straggler_tpcds_aws}, and~\ref{fig:token_bucket_CI})  causes behavior that breaks standard assumptions for statistical analysis (such as iid properties and stationarity). As an integral part of the experimentation procedure, samples collected should be tested for normality~\cite{shapirowilk}, independence~\cite{mann1947test}, and stationarity~\cite{dickey1979distribution}. When results are not normally-distributed, non-parametric statistics can be used~\cite{gibbons2011nonparametric}. When performance is not stationary, results can be limited to time periods when stationarity holds, or repetitions can be run over longer time frames, different diurnal or calendar cycles, etc. Techniques like CONFIRM~\cite{maricq2018taming} can be used to test whether confidence intervals converge as expected. 

It can also be helpful to discretize performance evaluation into units of time, e.g., \emph{one hour}. Gathering median performance for each interval, and applying techniques such as CONFIRM~\cite{maricq2018taming} over large-numbers of gathered medians results in statistically significant and realistic performance data. Large time periods can smooth out noise, helping to reduce unrepresentative measurements.

We also find it helpful to `rest' the infrastructure and randomize experiment order. Because it is hard to tell what performance-relevant state may build up in the hidden parts of the underlying cloud infrastructure, experimenters must ensure that the infrastructure is in as `neutral' a state as possible at the beginning of every experiment. The most reliable way to do so is to create a fresh set of VMs for every experiment. When running many small experiments, this can be cost- or time-prohibitive: in these cases, adding delays between experiments run in the same VMs can help. Data used while gathering baseline runs can be used to determine the appropriate length (e.g., seconds or minutes) of these rests. In addition, randomizing experiment order~\cite{abedi2017conducting} is a useful technique for avoiding self-interference.

\textbf{F5.5: Network performance on clouds is largely a function of provider implementation and policies, which can change at any time.}

Experimenters cannot treat ``the cloud'' as an opaque entity; results are significantly impacted by platform details that may or may not be public, and that are subject to change. (Indeed, much of the behavior that we document in Sections~\ref{sec:netw_var} and~\ref{sec:postgigabit_bigdata} is unlikely to be static over time.) Cloud experimenters can safeguard against this, somewhat, by publishing as much detail as possible about experiment setup (e.g., instance type, region, dates when experiments were run), establishing baseline performance numbers for the cloud itself, and only comparing results to future experiments when these baselines match.

\section{Related Work}
\label{sec:rel_work}

We have showed the extent of network performance variability in modern clouds, as well as how practitioners disregard cloud performance variability when designing and running experiments. Moreover, we have showed what the impact of network performance variability is on experiment design and on the performance of big data applications. 
We discuss our contributions in contrast to several categories of related work.

\textbf{Sound Experimentation (in the Cloud).} Several articles already discuss pitfalls of systems experiment design and presentation. Such work fits two categories: guidelines for better experiment design~\cite{abedi2017conducting,maricq2018taming,Kalibera2013Rigorous,Curtsinger2013Stabilizer} and avoiding logical fallacies in reasoning and presentation of empirical results~\cite{hoefler2015scientific, blackburn2016truth, oliveira2013quantile}. Adding to this type of work, we survey how practitioners apply such knowledge, and assess the impact of poor experiment design on the reliability of the achieved results. We investigate the impact of \emph{variability} on performance reproducibility, and uncover variability behavior on modern clouds.  

\textbf{Network Variability and Guarantees.} Network variability has been studied throughout the years in multiple contexts, such as HPC~\cite{bhatele2015identifying,bhatele2013there}, experimental testbeds~\cite{maricq2018taming} and virtualized environments~\cite{iosup2010performance, kossmann2010evaluation,schad2010runtime}. In the latter scenario, many studies have already assessed the performance variability of cloud datacenter networks~\cite{persico2015measuring,li2010cloudcmp,wang2010impact}. To counteract this behavior, cloud providers tackle the variability problem at the infrastructure level~\cite{raghavan2007cloud,conex}. In general, these approaches introduce network virtualization~\cite{guo2010secondnet,rodrigues2011gatekeeper}, or traffic shaping mechanisms~\cite{dalton2018andromeda}, such as the token buckets we identified, at the networking layer (per VM or network device), as well as a scheduling (and placement) policy framework~\cite{lacurts2013choreo}. 
In this work, we considered both types of variability: the one given by resource sharing and the one introduced by the interaction between applications and cloud QoS policies.


\textbf{Variability-aware Network Modeling, Simulation, and Emulation.}  
Modeling variable networks~\cite{madireddy2018modeling,gong2014finding} is a topic of interest. Kanev et al.~\cite{kanev2015profiling} profiled and measured more than 20,000 Google machines to understand the impact of performance variability on commonly used workloads in clouds. 
Uta et al. emulate gigabit real-world cloud networks to study their impact on the performance of batch-processing applications~\cite{uta2018performance}.
Casale and Tribastone \cite{casale2013modelling} model the exogenous variability of cloud workloads as continuous-time Markov chains. Such work cannot isolate the behavior of network-level variability compared to other types of resources. 

\section{Conclusion}

We studied the impact of cloud network performance variability, characterizing its impact on big data experiment reproducibility. We found that many articles disregard network variability in the cloud and perform a limited number of repetitions, which poses a serious threat to the validity of conclusions drawn from such experiment designs. We uncovered and characterized the network variability of modern cloud networks and showed that network performance variability leads to variable slowdowns and poor performance predictability, resulting in non-reproducible performance evaluations. To counter such behavior, we proposed protocols to achieve reliable cloud-based experimentation. As future work, we hope to extend this analysis to application domains other than big data and develop software tools to help experimenters run reproducible experiments in the cloud.



\section*{Code and Data Artifacts}

\textbf{Raw Cloud Data:}

\href{https://zenodo.org/badge/latestdoi/228254892}{DOI:10.5281/zenodo.3576604}

\noindent\textbf{Bandwidth Emulator:}

\href{https://github.com/alexandru-uta/bandwidth_emulator}{github.com/alexandru-uta/bandwidth\_emulator}

\noindent\textbf{Cloud Benchmarking:}

\href{https://github.com/alexandru-uta/measure-tcp-latency}{github.com/alexandru-uta/measure-tcp-latency}

\section*{Acknowledgements}

We thank all the anonymous reviewers for all their valuable suggestions, which helped improve this manuscript. Work on this article was funded via NWO VIDI MagnaData (\#14826), SURFsara e-infra180061, as well as NSF Grant numbers 1419199 and 1743363, and NSF OAC-1836650, NSF CNS-1764102, NSF CNS-1705021, NSF OAC-1450488, and the Center for Research in Open Source Software.

\bibliographystyle{abbrv}
\bibliography{references}

\begin{thebibliography}{10}

\bibitem{DBLP:conf/sosp/2017}
{\em Proceedings of the 26th Symposium on Operating Systems Principles,
  Shanghai, China, October 28-31, 2017}. {ACM}, 2017.

\bibitem{DBLP:conf/sc/2018}
{\em Proceedings of the International Conference for High Performance
  Computing, Networking, Storage, and Analysis, {SC} 2018, Dallas, TX, USA,
  November 11-16, 2018}. {IEEE} / {ACM}, 2018.

\bibitem{abedi2017conducting}
A.~Abedi and T.~Brecht.
\newblock Conducting repeatable experiments in highly variable cloud computing
  environments.
\newblock In {\em Proceedings of the 8th ACM/SPEC on International Conference
  on Performance Engineering}, pages 287--292. ACM, 2017.

\bibitem{armbrust2015spark}
M.~Armbrust, R.~S. Xin, C.~Lian, Y.~Huai, D.~Liu, J.~K. Bradley, X.~Meng,
  T.~Kaftan, M.~J. Franklin, A.~Ghodsi, et~al.
\newblock Spark sql: Relational data processing in spark.
\newblock In {\em Proceedings of the 2015 ACM SIGMOD International Conference
  on Management of Data}, pages 1383--1394. ACM, 2015.

\bibitem{DBLP:conf/osdi/2018}
A.~C. Arpaci{-}Dusseau and G.~Voelker, editors.
\newblock {\em 13th {USENIX} Symposium on Operating Systems Design and
  Implementation, {OSDI} 2018, Carlsbad, CA, USA, October 8-10, 2018}. {USENIX}
  Association, 2018.

\bibitem{aws_enhanced_networking}
{AWS Enhanced Networking}.
\newblock \url{https://aws.amazon.com/ec2/features/ \#enhanced-networking},
  2019.

\bibitem{ballani2011towards}
H.~Ballani, P.~Costa, T.~Karagiannis, and A.~Rowstron.
\newblock Towards predictable datacenter networks.
\newblock In {\em ACM SIGCOMM Computer Communication Review}, volume~41, pages
  242--253. ACM, 2011.

\bibitem{bhatele2013there}
A.~Bhatele, K.~Mohror, S.~H. Langer, and K.~E. Isaacs.
\newblock There goes the neighborhood: performance degradation due to nearby
  jobs.
\newblock In {\em Proceedings of the International Conference on High
  Performance Computing, Networking, Storage and Analysis}, page~41. ACM, 2013.

\bibitem{bhatele2015identifying}
A.~Bhatele, A.~R. Titus, J.~J. Thiagarajan, N.~Jain, T.~Gamblin, P.-T. Bremer,
  M.~Schulz, and L.~V. Kale.
\newblock Identifying the culprits behind network congestion.
\newblock In {\em 2015 IEEE International Parallel and Distributed Processing
  Symposium}, pages 113--122. IEEE, 2015.

\bibitem{blackburn2016truth}
S.~M. Blackburn, A.~Diwan, M.~Hauswirth, P.~F. Sweeney, J.~N. Amaral,
  T.~Brecht, L.~Bulej, C.~Click, L.~Eeckhout, S.~Fischmeister, et~al.
\newblock The truth, the whole truth, and nothing but the truth: A pragmatic
  guide to assessing empirical evaluations.
\newblock {\em ACM Transactions on Programming Languages and Systems (TOPLAS)},
  38(4):15, 2016.

\bibitem{Boudec2011Performance}
J.-Y.~L. Boudec.
\newblock {\em Performance Evaluation of Computer and Communication Systems}.
\newblock EFPL Press, 2011.

\bibitem{conex}
B.~Briscoe and M.~Sridharan.
\newblock Network performance isolation in data centres using congestion
  exposure (conex).
\newblock 2012.

\bibitem{cao2017performance}
Z.~Cao, V.~Tarasov, H.~P. Raman, D.~Hildebrand, and E.~Zadok.
\newblock On the performance variation in modern storage stacks.
\newblock In {\em 15th {USENIX} Conference on File and Storage Technologies
  ({FAST} 17)}, pages 329--344, 2017.

\bibitem{casale2013modelling}
G.~Casale and M.~Tribastone.
\newblock Modelling exogenous variability in cloud deployments.
\newblock {\em ACM SIGMETRICS Performance Evaluation Review}, 2013.

\bibitem{chaimov2016scaling}
N.~Chaimov, A.~Malony, S.~Canon, C.~Iancu, K.~Z. Ibrahim, and J.~Srinivasan.
\newblock Scaling spark on hpc systems.
\newblock In {\em Proceedings of the 25th ACM International Symposium on
  High-Performance Parallel and Distributed Computing}, pages 97--110. ACM,
  2016.

\bibitem{cohen1960coefficient}
J.~Cohen.
\newblock A coefficient of agreement for nominal scales.
\newblock {\em Educational and psychological measurement}, 20(1):37--46, 1960.

\bibitem{Curtsinger2013Stabilizer}
C.~Curtsinger and E.~D. Berger.
\newblock Stabilizer: Statistically sound performance evaluation.
\newblock {\em SIGARCH Comput. Archit. News}, 41(1):219--228, Mar. 2013.

\bibitem{dalton2018andromeda}
M.~Dalton, D.~Schultz, J.~Adriaens, A.~Arefin, A.~Gupta, B.~Fahs,
  D.~Rubinstein, E.~C. Zermeno, E.~Rubow, J.~A. Docauer, et~al.
\newblock Andromeda: Performance, isolation, and velocity at scale in cloud
  network virtualization.
\newblock In {\em 15th {USENIX} Symposium on Networked Systems Design and
  Implementation ({NSDI}'18)}, pages 373--387, 2018.

\bibitem{databricks_instances}
{Databricks Instance Types}.
\newblock \url{https://databricks.com/product/aws-pricing/instance-types},
  2019.

\bibitem{oliveira2013quantile}
A.~B. De~Oliveira, S.~Fischmeister, A.~Diwan, M.~Hauswirth, and P.~F. Sweeney.
\newblock Why you should care about quantile regression.
\newblock In {\em ACM SIGPLAN Notices}, volume~48, pages 207--218. ACM, 2013.

\bibitem{dean2013tail}
J.~Dean and L.~A. Barroso.
\newblock The tail at scale.
\newblock {\em Communications of the ACM}, 56(2):74--80, 2013.

\bibitem{dickey1979distribution}
D.~A. Dickey and W.~A. Fuller.
\newblock Distribution of the estimators for autoregressive time series with a
  unit root.
\newblock {\em Journal of the American Statistical Association},
  74(366a):427--431, 1979.

\bibitem{farley2012more}
B.~Farley, A.~Juels, V.~Varadarajan, T.~Ristenpart, K.~D. Bowers, and M.~M.
  Swift.
\newblock More for your money: exploiting performance heterogeneity in public
  clouds.
\newblock In {\em Proceedings of the Third ACM Symposium on Cloud Computing},
  page~20. ACM, 2012.

\bibitem{ghit2015reducing}
B.~Ghit and D.~Epema.
\newblock Reducing job slowdown variability for data-intensive workloads.
\newblock In {\em 2015 IEEE 23rd International Symposium on Modeling, Analysis,
  and Simulation of Computer and Telecommunication Systems}. IEEE, 2015.

\bibitem{gibbons2011nonparametric}
J.~D. Gibbons and S.~Chakraborti.
\newblock {\em Nonparametric statistical inference}.
\newblock Springer, 2011.

\bibitem{gong2014finding}
Y.~Gong, B.~He, and D.~Li.
\newblock Finding constant from change: Revisiting network performance aware
  optimizations on iaas clouds.
\newblock In {\em Proceedings of the International Conference for High
  Performance Computing, Networking, Storage and Analysis}, pages 982--993.
  IEEE Press, 2014.

\bibitem{google_networking}
{Google Andromeda Networking}.
\newblock
  \url{{https://cloud.google.com/blog/products/networking/google-cloud-networking-in-depth-how-andromeda-2-2-enables-high-throughput-vms}},
  2019.

\bibitem{grosvenor2015queues}
M.~P. Grosvenor, M.~Schwarzkopf, I.~Gog, R.~N. Watson, A.~W. Moore, S.~Hand,
  and J.~Crowcroft.
\newblock Queues don’t matter when you can {JUMP} them!
\newblock In {\em 12th {USENIX} Symposium on Networked Systems Design and
  Implementation ({NSDI}'15)}, pages 1--14, 2015.

\bibitem{guo2010secondnet}
C.~Guo, G.~Lu, H.~J. Wang, S.~Yang, C.~Kong, P.~Sun, W.~Wu, and Y.~Zhang.
\newblock Secondnet: a data center network virtualization architecture with
  bandwidth guarantees.
\newblock In {\em Proceedings of the 6th International COnference}, page~15.
  ACM, 2010.

\bibitem{hoefler2015scientific}
T.~Hoefler and R.~Belli.
\newblock Scientific benchmarking of parallel computing systems: twelve ways to
  tell the masses when reporting performance results.
\newblock In {\em Proceedings of the international conference for high
  performance computing, networking, storage and analysis}. ACM, 2015.

\bibitem{huang2010hibench}
S.~Huang, J.~Huang, J.~Dai, T.~Xie, and B.~Huang.
\newblock The hibench benchmark suite: Characterization of the mapreduce-based
  data analysis.
\newblock In {\em Data Engineering Workshops (ICDEW), 2010 IEEE 26th
  International Conference on}, pages 41--51. IEEE, 2010.

\bibitem{hubert2002linux}
B.~Hubert, T.~Graf, G.~Maxwell, R.~van Mook, M.~van Oosterhout, P.~Schroeder,
  J.~Spaans, and P.~Larroy.
\newblock Linux advanced routing \& traffic control.
\newblock In {\em Ottawa Linux Symposium}, page 213, 2002.

\bibitem{iosup2018massivizing}
A.~Iosup, A.~Uta, L.~Versluis, G.~Andreadis, E.~Van~Eyk, T.~Hegeman,
  S.~Talluri, V.~Van~Beek, and L.~Toader.
\newblock Massivizing computer systems: a vision to understand, design, and
  engineer computer ecosystems through and beyond modern distributed systems.
\newblock In {\em 2018 IEEE 38th International Conference on Distributed
  Computing Systems (ICDCS)}, pages 1224--1237. IEEE, 2018.

\bibitem{iosup2010performance}
A.~Iosup, N.~Yigitbasi, and D.~Epema.
\newblock On the performance variability of production cloud services.
\newblock In {\em Cluster, Cloud and Grid Computing (CCGrid), 2011 11th
  IEEE/ACM International Symposium on}, pages 104--113. IEEE, 2011.

\bibitem{jain1990art}
R.~Jain.
\newblock {\em The art of computer systems performance analysis: techniques for
  experimental design, measurement, simulation, and modeling}.
\newblock John Wiley \& Sons, 1990.

\bibitem{jimenez2017popper}
I.~Jimenez, M.~Sevilla, N.~Watkins, C.~Maltzahn, J.~Lofstead, K.~Mohror,
  A.~Arpaci-Dusseau, and R.~Arpaci-Dusseau.
\newblock The popper convention: Making reproducible systems evaluation
  practical.
\newblock In {\em 2017 IEEE International Parallel and Distributed Processing
  Symposium Workshops (IPDPSW)}, pages 1561--1570. IEEE, 2017.

\bibitem{Kalibera2013Rigorous}
T.~Kalibera and R.~Jones.
\newblock Rigorous benchmarking in reasonable time.
\newblock {\em SIGPLAN Not.}, 48, 2013.

\bibitem{kanev2015profiling}
S.~Kanev, J.~P. Darago, K.~Hazelwood, P.~Ranganathan, T.~Moseley, G.-Y. Wei,
  and D.~Brooks.
\newblock Profiling a warehouse-scale computer.
\newblock In {\em ACM SIGARCH Computer Architecture News}, volume~43, pages
  158--169. ACM, 2015.

\bibitem{kossmann2010evaluation}
D.~Kossmann, T.~Kraska, and S.~Loesing.
\newblock An evaluation of alternative architectures for transaction processing
  in the cloud.
\newblock In {\em Proceedings of the 2010 ACM SIGMOD International Conference
  on Management of data}. ACM, 2010.

\bibitem{lacurts2013choreo}
K.~LaCurts, S.~Deng, A.~Goyal, and H.~Balakrishnan.
\newblock Choreo: Network-aware task placement for cloud applications.
\newblock In {\em Proceedings of the 2013 conference on Internet measurement
  conference}, pages 191--204. ACM, 2013.

\bibitem{leitner2016patterns}
P.~Leitner and J.~Cito.
\newblock Patterns in the chaos--a study of performance variation and
  predictability in public iaas clouds.
\newblock {\em ACM Transactions on Internet Technology (TOIT)}, 16(3):15, 2016.

\bibitem{li2010cloudcmp}
A.~Li, X.~Yang, S.~Kandula, and M.~Zhang.
\newblock Cloudcmp: comparing public cloud providers.
\newblock In {\em Proceedings of the 10th ACM SIGCOMM conference on Internet
  measurement}, pages 1--14. ACM, 2010.

\bibitem{DBLP:conf/nsdi/2019}
J.~R. Lorch and M.~Yu, editors.
\newblock {\em 16th {USENIX} Symposium on Networked Systems Design and
  Implementation, {NSDI} 2019, Boston, MA, February 26-28, 2019}. {USENIX}
  Association, 2019.

\bibitem{madireddy2018modeling}
S.~Madireddy, P.~Balaprakash, P.~Carns, R.~Latham, R.~Ross, S.~Snyder, and
  S.~Wild.
\newblock Modeling i/o performance variability using conditional variational
  autoencoders.
\newblock In {\em 2018 IEEE International Conference on Cluster Computing
  (CLUSTER)}, pages 109--113. IEEE, 2018.

\bibitem{mann1947test}
H.~B. Mann and D.~R. Whitney.
\newblock On a test of whether one of two random variables is stochastically
  larger than the other.
\newblock {\em The Annals of Mathematical Statistics}, pages 50--60, 1947.

\bibitem{maricq2018taming}
A.~Maricq, D.~Duplyakin, I.~Jimenez, C.~Maltzahn, R.~Stutsman, and R.~Ricci.
\newblock Taming performance variability.
\newblock In {\em 13th {USENIX} Symposium on Operating Systems Design and
  Implementation ({OSDI} 18)}, pages 409--425, 2018.

\bibitem{mogul2012we}
J.~C. Mogul and L.~Popa.
\newblock What we talk about when we talk about cloud network performance.
\newblock {\em ACM SIGCOMM Computer Communication Review}, 42(5):44--48, 2012.

\bibitem{nambiar2006making}
R.~O. Nambiar and M.~Poess.
\newblock The making of tpc-ds.
\newblock In {\em VLDB}, 2006.

\bibitem{ousterhout2015making}
K.~Ousterhout, R.~Rasti, S.~Ratnasamy, S.~Shenker, and B.-G. Chun.
\newblock Making sense of performance in data analytics frameworks.
\newblock In {\em {NSDI} '15}, volume~15, pages 293--307, 2015.

\bibitem{persico2015measuring}
V.~Persico, P.~Marchetta, A.~Botta, and A.~Pescap{\'e}.
\newblock Measuring network throughput in the cloud: the case of amazon ec2.
\newblock {\em Computer Networks}, 93:408--422, 2015.

\bibitem{raghavan2007cloud}
B.~Raghavan, K.~Vishwanath, S.~Ramabhadran, K.~Yocum, and A.~C. Snoeren.
\newblock Cloud control with distributed rate limiting.
\newblock {\em ACM SIGCOMM Computer Communication Review}, 37(4):337--348,
  2007.

\bibitem{rodrigues2011gatekeeper}
H.~Rodrigues, J.~R. Santos, Y.~Turner, P.~Soares, and D.~O. Guedes.
\newblock Gatekeeper: Supporting bandwidth guarantees for multi-tenant
  datacenter networks.
\newblock In {\em WIOV}, 2011.

\bibitem{schad2010runtime}
J.~Schad, J.~Dittrich, and J.-A. Quian{\'e}-Ruiz.
\newblock Runtime measurements in the cloud: observing, analyzing, and reducing
  variance.
\newblock {\em Proceedings of the VLDB Endowment}, 3(1-2):460--471, 2010.

\bibitem{shapirowilk}
S.~Shapiro and M.~Wilk.
\newblock An analysis of variance test for normality (complete samples).
\newblock {\em Biometricka}, 52:591--611, Dec. 1965.

\bibitem{suresh2015c3}
L.~Suresh, M.~Canini, S.~Schmid, and A.~Feldmann.
\newblock C3: Cutting tail latency in cloud data stores via adaptive replica
  selection.
\newblock In {\em 12th {USENIX} Symposium on Networked Systems Design and
  Implementation ({NSDI}'15)}, pages 513--527, 2015.

\bibitem{trivedi2018albis}
A.~Trivedi, P.~Stuedi, J.~Pfefferle, A.~Schuepbach, and B.~Metzler.
\newblock Albis: High-performance file format for big data systems.
\newblock In {\em 2018 {USENIX} Annual Technical Conference ({USENIX ATC} 18)},
  pages 615--630, 2018.

\bibitem{our_traces}
A.~Uta, A.~Custura, D.~Duplyakin, I.~Jimenez, J.~Rellermeyer, C.~Maltzahn,
  R.~Ricci, and A.~Iosup.
\newblock {Cloud Network Performance Variability Repository}.
\newblock \url{https://zenodo.org/record/3576604#.XfeXaOuxXOQ}, 2019.

\bibitem{uta2018performance}
A.~Uta and H.~Obaseki.
\newblock A performance study of big data workloads in cloud datacenters with
  network variability.
\newblock In {\em Companion of the 2018 ACM/SPEC International Conference on
  Performance Engineering}. ACM, 2018.

\bibitem{viera2005understanding}
A.~J. Viera, J.~M. Garrett, et~al.
\newblock Understanding interobserver agreement: the kappa statistic.
\newblock {\em Fam med}, 37(5):360--363, 2005.

\bibitem{wang2017using}
C.~Wang, B.~Urgaonkar, N.~Nasiriani, and G.~Kesidis.
\newblock Using burstable instances in the public cloud: Why, when and how?
\newblock {\em Proceedings of the ACM on Measurement and Analysis of Computing
  Systems}, 1(1):11, 2017.

\bibitem{wang2010impact}
G.~Wang and T.~E. Ng.
\newblock The impact of virtualization on network performance of amazon ec2
  data center.
\newblock In {\em INFOCOM}. IEEE, 2010.

\bibitem{white2012hadoop}
T.~White.
\newblock {\em Hadoop: The definitive guide}.
\newblock " O'Reilly Media, Inc.", 2012.

\bibitem{zaharia2012resilient}
M.~Zaharia, M.~Chowdhury, T.~Das, A.~Dave, J.~Ma, M.~McCauley, M.~J. Franklin,
  S.~Shenker, and I.~Stoica.
\newblock Resilient distributed datasets: A fault-tolerant abstraction for
  in-memory cluster computing.
\newblock In {\em {NSDI}'12}. USENIX, 2012.

\end{thebibliography}


%
\end{document}